\documentclass[10pt,epsf,letterpaper]{article}

\usepackage{amsmath}
\usepackage{amsfonts}
\usepackage{verbatim}
\usepackage{amssymb}
\usepackage{graphicx}
\usepackage{mathrsfs}
\usepackage{physics}
\usepackage{hyperref}
\usepackage[utf8]{inputenc} 
\usepackage{eurosym}
\usepackage{physics}
\usepackage[svgnames]{xcolor} 
\usepackage{amsmath,amssymb,amsfonts,amsthm}
\usepackage{mathtools,old-arrows,verbatim,mathrsfs}
\usepackage{graphicx}
\usepackage{epstopdf,epsfig,changepage}

\usepackage{bm}
\usepackage[top=2.5cm, bottom=3cm, left=3.5cm, right=3.5cm,
heightrounded,marginparwidth=1.5cm, marginparsep=1cm]{geometry}
\usepackage[shortlabels]{enumitem}
\usepackage[square,numbers,merge,comma,sort&compress]{natbib} 
\makeatletter
\def\NAT@spacechar{\,}  
\makeatother
\usepackage{comment}
\usepackage{relsize,setspace,moresize}
\usepackage{latexsym,mathrsfs,calligra,aurical}
\usepackage{xargs,extarrows,empheq,url}

\usepackage[version=3]{mhchem}

\usepackage{tikz}
\usetikzlibrary{arrows}
\hypersetup{colorlinks=true,	linkcolor=blue,	filecolor=magenta, 	urlcolor=blue,	citecolor=blue,	pdftitle={Exact hairy black holes asymtotically $AdS2+1$}, pdfpagemode=FullScreen}
\usepackage{footnotebackref}
\usepackage{hypernat}

\DeclareFixedFont\trfont{OT1}{phv}{b}{sc}{11}
\def\={\:=\:}

\newcommand{\br}{\biggr}
\newcommand{\bl}{\biggl}

\usepackage{ulem}

\providecommand{\U}[1]{\protect\rule{.1in}{.1in}}

\def\Om{\Omega}

\graphicspath{{./fig_crop/}}
\renewcommand{\(}{\left(}
\renewcommand{\)}{\right)}
\renewcommand{\[}{\left[}
\renewcommand{\]}{\right]}
\def\Om{\Omega}
\newcommandx{\ETh}[2][1=M,2=\alpha,usedefault]{\Theta_{#1}{}^{#2}}
\newcommandx{\overbar}[1]{\mkern1.5mu\overline{\mkern-2.0mu#1\mkern-2.0mu}\mkern1.5mu}
\newcommandx{\overbarM}[1]{\mkern6.0mu\overline{\mkern-5.5mu#1\mkern-3.5mu}\mkern1.5mu}
\newcommandx{\overbarcal}[1]{\mkern6.0mu\overline{\mkern-5.5mu#1\mkern-1.0mu}\mkern1.5mu}
\DeclareFixedFont\trfont{OT1}{phv}{b}{sc}{11}
\DeclareMathAlphabet{\mathpzc}{OT1}{pzc}{m}{it}
\DeclareMathAlphabet{\mathcal}{OMS}{cmsy}{m}{n}
\DeclareSymbolFontAlphabet{\Scr}{rsfs}
\DeclareMathAlphabet{\mathbold}{U}{BOONDOX-ds}{m}{n}
\SetMathAlphabet{\mathbold}{bold}{U}{BOONDOX-ds}{b}{n}
\DeclareMathAlphabet{\mathcalboondox}{U}{BOONDOX-calo}{m}{n}
\SetMathAlphabet{\mathcalboondox}{bold}{U}{BOONDOX-calo}{b}{n}
\DeclareMathAlphabet{\mathbcalboondox}{U}{BOONDOX-calo}{b}{n}

\usepackage{subcaption} 
\usepackage[title]{appendix}
\usepackage{multirow}
\usepackage[numbers]{natbib} 
\usepackage{titlesec}
\usepackage{etoolbox, chngcntr}
\AtBeginEnvironment{appendices}{%
 \titleformat{\section}{\bfseries\Large}{\appendixname~\thesection:}{0.5em}{}%
 \titleformat{\subsection}{\bfseries\large}{\thesubsection}{0.5em}{}%
\counterwithin{equation}{section}
}
\usepackage{floatrow}

\begin{document}


\title{\centering\boldmath\LARGE\bfseries{%
	Exact hairy black holes asymptotically $AdS_{2+1}$
	}\vspace{1.25em}}

\author{Carlos Desa$^{(1)}$, Weyner Ccuiro$^{(1)}$ and David Choque$^{(1)}$ \\
\\\textit{$^{(1)}$Universidad Nacional de San Antonio Abad del Cusco,} \\\textit{Av. La Cultura 733, Cusco, Per\'u.}}

\maketitle
\begin{abstract}
 In the context of Einstein's minimally coupled scalar field theory, we present a family of hairy black holes which are asymptotically $AdS_{2+1}$. We investigate the boundary conditions and build the thermal superpotential. Two methods are used to regularize the free energy in the Euclidean section and the Brown-Newton tensor in the Lorentzian section. Finally, the relevant thermodynamic quantities are calculated and the different phases are analyzed.
\end{abstract}
{\hypersetup{linkcolor=black}
\tableofcontents}

\newpage

\section{Introduction}
%
In accordance with the no-hair theorem \cite{Heusler:1992ss,Nunez:1996xv}, in three dimensions there are no asymptotically flat black hole solutions for pure gravity. Indeed, the Riemann tensor is purely a function of the Ricci tensor and metric $R_{\alpha\beta\gamma\sigma}(R_{\mu\nu},g_{\mu\nu})$ on this manifold. Then, no solutions emerge unless the Einstein-Hilbert action is supplemented with a negative cosmological constant, as considered by Ba\~nados, Teitelboim and Zanelli (BTZ), who first found the  BTZ black hole solution \cite{Banados:1992wn,Banados:1992gq}, which have been extensively studied \cite{Carlip:1995qv,Strominger:1997eq,Balthazar:2021xeh,Anninos:2008fx,Peet:2000hn,Hartman:2014oaa}. The procedure for overcome the no-hair theorem
consists the coupling the other fields to Einstein-Hilbert action minimally, non-minimally or conformally \cite{Correa:2011dt,Xu:2014uka,Zhao:2013isa,Perez:2012cf,Bueno:2021krl,Nunez:1996xv,Greene:1992fw}. A further way to circumvent the no-hair theorem is by considering higher-derivative gravities \cite{Bueno:2017sui,Bueno:2022lhf,Cano:2022wwo}, this has been successfully done in $D=3+1$ dimensions \cite{Martinez:2005di,Astefanesei:2019qsg, Anabalon:2022ksf, Anabalon:2020pez,Gonzalez:2013aca,Martinez:2004nb,Martinez:2002ru}.

Here we show the construction of the hairy family black hole solution. We begin by integrating the metric functions, then the scalar field $\phi$, and finally the scalar potential $V(\phi)$. This method is the on-shell procedure, and it was extensively applied in \cite{Anabalon:2013sra,Anabalon:2013qua,Acena:2012mr,Acena:2013jya}. 
In summary, we present a family of hairy black hole solutions in three dimensions with a non-trivial scalar potential minimally coupling. This scalar field contributes to the metric and, therefore, to the thermodynamics of the black hole. In \cite{Hawking:1982dh} was discovered
the existence of the first order Hawking-Page phase transitions and their respective dual interpretations \cite{Witten:1998zw}, based on the conjecture $AdS/CFT$ \cite{Aharony:1999ti,Gubser:1998bc,Witten:1998qj}, this led to a great deal of activity in research on black hole phase transitions \cite{Anabalon:2019tcy,Anabalon:2015ija,Chamblin:1999hg,Lu:2013ura,Giribet:2014fla}. It is known that the scalar hair can have relevant effects on phase diagrams. Here \cite{Astefanesei:2020xvn, Anabalon:2019tcy,Astefanesei:2019mds} they have shown a window of parameters in which the asymptotically flat, hairy black holes are stable. These last results encourage us to study the phase diagrams of our family of hairy black holes. In general, the most interesting phase diagrams can be found in the context of $VdP$ formalism \cite{Kubiznak:2012wp,Gunasekaran:2012dq,Astefanesei:2019ehu,Astefanesei:2020xvn}, but we focused on phase diagrams with a fixed cosmological constant. \cite{Anabalon:2015ija,Giribet:2014fla,Astefanesei:2019mds,Anabalon:2019tcy,Astefanesei:2018vga}\\

Recently, hairy solutions in $D=3+1$
was embed in omega-deformed gauged $N=8$
supergravity \cite{Anabalon:2013eaa},
and the explicit construction of the hairy solution from $N=2$ supergravity was demonstrated
\cite{Anabalon:2017yhv,Anabalon:2020qux}. These last results open the possibility of finding the supergravity origin of our solution.
Many solutions with different conformal masses can be found in the literature \cite{Henneaux:2002wm,Tang:2019jkn,Correa:2010hf}, in our case, we have an interesting example
with a conformal mass $m^{2}=-1/L^{2}$ that saturates the Breitenlohner-Freedman 
bound $m_{BF}^{2}=-1/L^{2}$.

The well-posed variational principle ensures the construction of the counterterms for the scalar field, allowing us to obtain the finite on-shell action\cite{Anabalon:2015xvl}. In the context of supergravity theories, the theorem of positive energy is ensured by the existence of a superpotential\cite{Hertog:2004dr,Gallerati:2021cty,Papadimitriou:2007sj},
curiously, in \cite{Batrachenko:2004fd} they proposed a superpotential as a counterterm; the problem is that in general, we cannot construct this superpotential exactly. The goal of \cite{Gursoy:2008za}
was to develop a thermal superpotential that considered the black hole horizon's existence.
This thermal superpotential is easier to construct, and similarly to \cite{Anabalon:2020qux,Astefanesei:2021ryn} we use it 
to regularize the on-shell action and the Brown-York quasilocal stress tensor \cite{Brown:1992br} \\

We structure this work in the following way.
Section \ref{sec:2} describes the theory's action, as well as the potential $V(\phi)$ and its stability conditions, which allow us to avoid the non-hair theorem. We show an exact hairy solution in the section \ref{soluADS}; we also consider the horizon existence conditions, the thermal superpotential, and the asymptotic $AdS$ boundary symmetry for the scalar field. In the middle part, section \ref{sec:4}, we get the finite on-shell action, and the goal is the construction of the counterterm for the scalar field. In the section \ref{sec:5} we build the quasilocal stress tensor properly regularized by two different methods. In the last two sections, \ref{sec:6} and \ref{sec:7} we study the thermodynamics, show the conclusions, and discuss future directions.\\


\newpage
%
\section{Theory}
\label{sec:2}
%
In $D=2+1$ dimensions, we are interested in modified Einstein-Hilbert actions with a non-minimal coupling scalar field. 
\begin{equation}
I[g_{\mu\nu},\phi]=\frac{1}{2\kappa}\int_{\mathcal{M}}{d^{3}x\sqrt{-g}\biggl{[}R-\frac{(\partial\phi)^{2}}{2}-V(\phi)\biggr{]}}+\frac{1}{\kappa}%
\int_{\partial\mathcal{M}}{d^{2}xK\sqrt{-h}}+I_{\phi}%
    \label{action1}
\end{equation}
where $V(\phi)$ is the scalar potential, $\kappa=8\pi G_{N}$, and
the last term is the Gibbons-Hawking boundary term. The boundary metric is $h_ab$, and the trace of the extrinsic curvature is $K$. Then the equations of motion for the metric are
\begin{equation}
    E_{\alpha\beta}=G_{\alpha\beta}-T_{\alpha\beta}=0
\end{equation}
\begin{equation}
    T_{\alpha\beta}=\frac{1}{2}\partial_{\alpha}\phi\partial_{\beta}\phi-\frac{g_{\alpha\beta}}{2}\[\frac{(\partial\phi)^{2}}{2}+V(\phi)\]
\end{equation}
where the Einstein's tensor is $G_{\alpha\beta}:=R_{\alpha\beta}-Rg_{\alpha\beta}/2$.
And the Klein-Gordon equation for the scalar field is
%
\begin{equation}
    \frac{1}{\sqrt{-g}}\partial_{\alpha}\(\sqrt{-g}g^{\alpha\beta}\partial_{\beta}\phi\)-\frac{\partial V}{\partial\phi}=0
\end{equation}
\section{Asymptotically $AdS$ solution}
\label{soluADS}
\begin{figure}[ht]
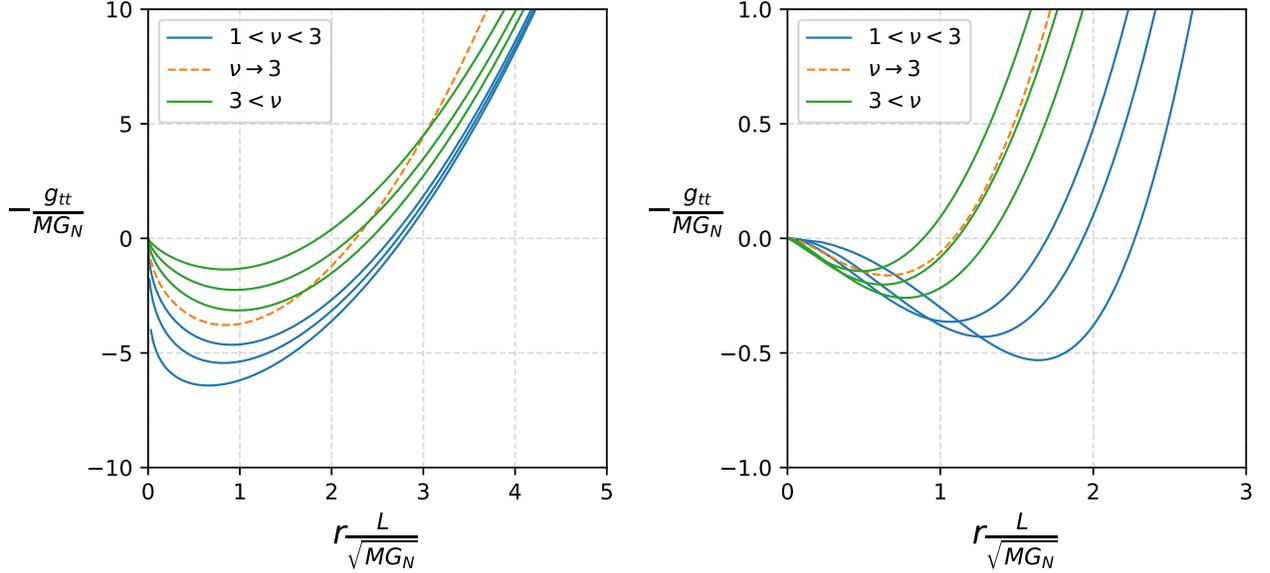

    \centering
    \includegraphics[width=0.5\linewidth]{pbgttvsr.jpg}\includegraphics[width=0.5\linewidth]{nbgttvsr.jpg}
    \caption{One can study the event horizon of a hairy black hole by analyzing the roots of the temporal metric component $-g_{tt}(x)=0$.
    $-g_{tt}=f(x)\Omega(x)$ versus radial coordinate is plotted. We fix $\alpha=9/L^2$ and $~r(x)=\sqrt{\Omega(x)}$. Also, we can deduce two important conclusions from these pictures for each branch:
    At low values of the hairy parameter $1<\nu<3$, the event horizon is increased.
    And the event horizons for the positive branch (left picture) are smaller than those in the negative branch (right picture).
    }
    \label{fig:2}
\end{figure}

%
We consider the $D=2+1$ version of the metric  which is static and radially symmetric \cite{Anabalon:2012ta}. The non-dimensional radial coordinate is $x$ and we will show that we have two branches, $x\in(0,1)$ and $x\in(1,\infty)$. Each has a distinct singularity at $x=0$ and $x=\infty$, but they share a boundary at $x=1$
\begin{equation}
\label{xcoor}
    ds^{2}=\Omega(x)\left[  -f(x)dt^{2}+\frac{\eta^{2}dx^{2}}{f(x)}+d\varphi^{2}\right]
\end{equation}
Einstein's equations are
\begin{align}
\label{einsxx}
     & E_{t}^{~t}-E_{x}^{~x}=0 ~~\Rightarrow~~ (\phi^{\prime})^{2}=\frac{3\Omega^{\prime 2}-2\Omega\Omega^{\prime\prime}}{2\Omega^{2}} \\ \nonumber
     & E_{t}^{~t}-E_{\varphi}^{~\varphi}=0 ~~\Rightarrow~~ (\Omega^{1/2}f^{\prime})^{\prime}=0                                         \\ \nonumber
     & E_{t}^{~t}+E_{\varphi}^{~\varphi}=0 ~~\Rightarrow~~ V=-\frac{1}{4\eta^{2}\Omega(x)^3}\[
        2\Om^{\prime\prime}f\Om+2f^{\prime\prime}\Om^{2}-f\Om^{\prime 2}+3\Om\Om^{\prime}f^{\prime}\]
\end{align}
The conformal factor $\Omega(x)$ was constructed in \cite{Anabalon:2013sra,Anabalon:2013qua,Acena:2012mr,Acena:2013jya}, and from it we can integrate $f(x)$ and $\phi(x)$. This conformal factor blowing up at the boundary
$x=1$, and $\eta$ is a constant integration related to the hairy black hole's mass
\begin{equation}
    \Omega(x)=\frac{\nu^{2}x^{\nu-1}}{\eta^{2}(x^{\nu}-1)^{2}}
    \label{om1}
\end{equation}
$\Omega(x)$ is the input function, and from $(\Omega^{1/2}f^{\prime})^{\prime}=0$ we can integrate $f(x)$, with new a constant $\alpha$.\footnote{Actually, $\alpha$ is not a constant integration. Considering $f(x)$ 
and $\Omega(x)$ in $E_{t}^{~t}+E_{\varphi}^{~\varphi}=0$
We obtain a scalar potential that is dependent on $\alpha$, and we can identify $\alpha$ as a constant of the theory. For details, see the next subsection \ref{Scpo}}.
There is no horizon when the theory parameter $\alpha$ is null, as shown in (\ref{metric11}); that is, solutions with $\alpha=0$ have a naked singularity.\footnote{We obtain the asymptotically flat solution at the limit $L\rightarrow\infty$; see the appendix \ref{apendiceA}}. Another property is that $f(\nu)=f(-\nu)$ and $\Omega(\nu)=\Omega(-\nu)$, so we can work with $\nu\geq 1$ without losing generality
\begin{equation}
    f(x)=\frac{1}{L^{2}}+\frac{\alpha}{2\nu}\bl{[}
    \frac{2\nu}{\nu^2-9} -\frac{1}{\(3+\nu\)}x^{\frac{3+\nu}{2}}+\frac{1}{\(3-\nu\)}x^{\frac{3-\nu}{2}}\br{]}
    \label{metric11}
\end{equation}
The scalar field $\phi(x)$ has the simple form (\ref{sca1}), where $\ell$ is the dilaton length. Clearly, if $\nu=1$, the non-hair limit is obtained, and the scalar field is $\phi(x)=0$. The scalar field is null $\phi=0$ at the boundary $x=1$, and it is blowing up near the singularity $x=0$ or $x=\infty$ 
\begin{equation}
    \phi(x)=\ell^{-1}\ln{x}, \qquad \ell^{-1}=\frac{\sqrt{2}}{2}\sqrt{\nu^{2}-1}
    \label{sca1}
\end{equation}
The existence of the horizon $f(x_{h},\alpha,\nu)=0$. is fixed by the correct choice of the theory's parameters $(\alpha,\nu)$, which are related to the convexity condition on $V(\phi)$,\footnote{The on-shell scalar potential $V(\phi)$ will be constructed in the following subsection \ref{Scpo}} i.e. the non-hair theorem; see Figure (\ref{fig:1}). We are going to study the existing conditions on the horizon for each branch; see Figure (\ref{fig:2}):
\subsection*{Negative branch}
\label{NB}
The scalar field is definite negative $\phi\leq 0$ in the region $0\leq x<1$, with the singularity at $x=0$ and the boundary at $x=1$
\begin{itemize}
    \item For: $\nu>3$ with $x\rightarrow 0$ the existence of the horizon $f(x_{h})=0$ is ensured by $f(x)<0$
\begin{equation}
 \lim_{x \to 0} f(x)\sim-\frac{\alpha L^{2}}{2\nu\qty(\nu-3)}x^{-\frac{\qty(\nu-3)}{2}}<0 \quad~\Rightarrow~ \alpha L^{2}>0 
\label{3.6}
\end{equation}

    \item For: $1\leq\nu<3$ with $x\rightarrow 0$ the existence of the horizon $f(x_{h})=0$ is ensured by $f(x)<0$

\begin{equation}
 \lim_{x \to 0}f(x)\sim\dfrac{1}{L^{2}}-\dfrac{\alpha}{9-\nu^{2}}<0 ~\quad~\Rightarrow~~ 0<9-\nu^{2}<\alpha L^{2}
 \label{3.7}
\end{equation}
\end{itemize}
\subsection*{Positive branch}
\label{PB}
The scalar field is definite positive, $\phi\geq 0$, in the region $1\leq x<\infty$, where the singularity is at $x=infty$ and the boundary is at $x=1$
\begin{itemize}
    \item The case $1\leq\nu<3$ and $\nu>3$ with $x\rightarrow \infty$ the horizon existence $f(x_{h})=0$ is ensured by $f(x)<0$ 
\begin{equation}
    \lim_{x \to \infty}f(x)\sim-\dfrac{\alpha}{2\nu\qty(3+\nu)}x^{\frac{3+\nu}{2}}<0~\quad \Rightarrow~ \alpha L^{2}>0 
    \label{3.8}
\end{equation}

\end{itemize}
%
\subsection{On-shell scalar potential}\label{Scpo}
%
In this section, we build our theory's on-shell scalar potential, $V(\phi)$. We use (\ref{om1}), (\ref{metric11}), and simplify to get $V(x)$ in the third equation (\ref{einsxx}), for more details, see the appendix \ref{apendiceAA}
\begin{multline}
\label{pot0}
V(x)  =-\frac{3}{4\nu^{2}}\qty(\frac{1}{L^{2}}+\frac{\alpha}{\nu^{2}-9})x^{-\frac{\sqrt{2\nu^{2}-2}}{2}}\bl{[}(1+\nu) \bl{(}1+\frac{\nu}{3}\br{)}x^{\frac{\sqrt{2\nu^{2}-2}}{2}}+
(1-\nu)\bl{(}1-\frac{\nu}{3}\br{)}x^{-\frac{\sqrt{2\nu^{2}-2}}{2}}\\
-2(1-\nu^{2})\br{]} -\frac{\alpha\cdot x^{\frac{\sqrt{2\nu^{2}-2}}{4}}}{\nu(\nu^2-9)}\[(1-\nu)x^{\frac{\sqrt{2\nu^{2}-2}}{4}}-(1+\nu)x^{-\frac{\sqrt{2\nu^{2}-2}}{4}}\]
\end{multline}
The correct form of the scalar potential is $V(\phi)$ rather than $V(x)$, and we get $x(\phi)$ from (\ref{sca1})
\begin{equation}
    \phi(x)=\ell^{-1}\ln{x} ~\Rightarrow~ x=e^{\ell\phi}
    \label{sca2}
\end{equation}
Finally, we obtain the theory's scalar potential $V(\phi)$ with parameters $\alpha$, $\Lambda=-2/L2$, and $\nu$. Another interesting construction was present in \cite{Anabalon:2013eaa,Anabalon:2017yhv,Anabalon:2012ih,Anabalon:2016izw}
\begin{multline}
\label{pot}
    V(\phi)  =-\frac{3\exp(-\phi\ell)}{4\nu^{2}}\qty(\frac{1}{L^{2}}+\frac{\alpha}{\nu^{2}-9})\bl{[}(1+\nu) \bl{(}1+\frac{\nu}{3}\br{)}\exp\(\phi\ell\nu\)+
    (1-\nu)\bl{(}1-\frac{\nu}{3}\br{)}\exp\(-\phi\ell\nu\)\\
    -2(1-\nu^{2})\br{]} -\frac{\alpha\exp(\phi\ell /2)}{\nu(\nu^2-9)}\[(1-\nu)\exp\(\frac{\phi\ell\nu}{2}\)-(1+\nu)\exp\(-\frac{\phi\ell\nu}{2}\)\]
\end{multline}
The self-interaction of the scalar potential lets the scalar field stay out of the horizon. From (\ref{sca1}) the scalar field at the boundary is null $\phi=0$, so the scalar potential near the boundary gives us the conformal mass $m^{2}=d^{2}V/d\phi^{2}\vert_{\phi=0}=-1/L^{2}$
\begin{equation}
    V(\phi)=-\frac{2}{L^{2}}-\frac{1}{2L^{2}}\phi^{2}-\frac{\ell^{3}(\nu^{2}-1)}{12L^{2}}\phi^{3}+\mathcal{O}(\phi^{4})~~\Rightarrow~~ V(0)<0
\end{equation}
At the same time, we can verify that the scalar potential $V(\phi)$ has a global maximum at the boundary, and thus $V(\phi)$ has a minimum at the horizon, ensuring the scalar field's global stability
\begin{equation}
    \frac{dV}{d\phi}=-\frac{1}{L^{2}}\phi-\frac{\ell^{3}(\nu^{2}-1)}{4L^{2}}\phi^{2}+\mathcal{O}(\phi^{3}) ~~~\Rightarrow~~~ \frac{dV}{d\phi}\br{\vert}_{\phi=0}=0
\end{equation}
\begin{equation}
    \frac{d^{2}V}{d\phi^{2}}=
    -\frac{1}{L^{2}}-\frac{\ell^{3}(\nu^{2}-1)}{2L^{2}}\phi+\mathcal{O}(\phi^{2})
    ~~~\Rightarrow~~~ \frac{d^{2}V}{d\phi^{2}}\br{\vert}_{\phi=0}<0
    \label{esca1}
\end{equation}
%
%
\begin{figure}
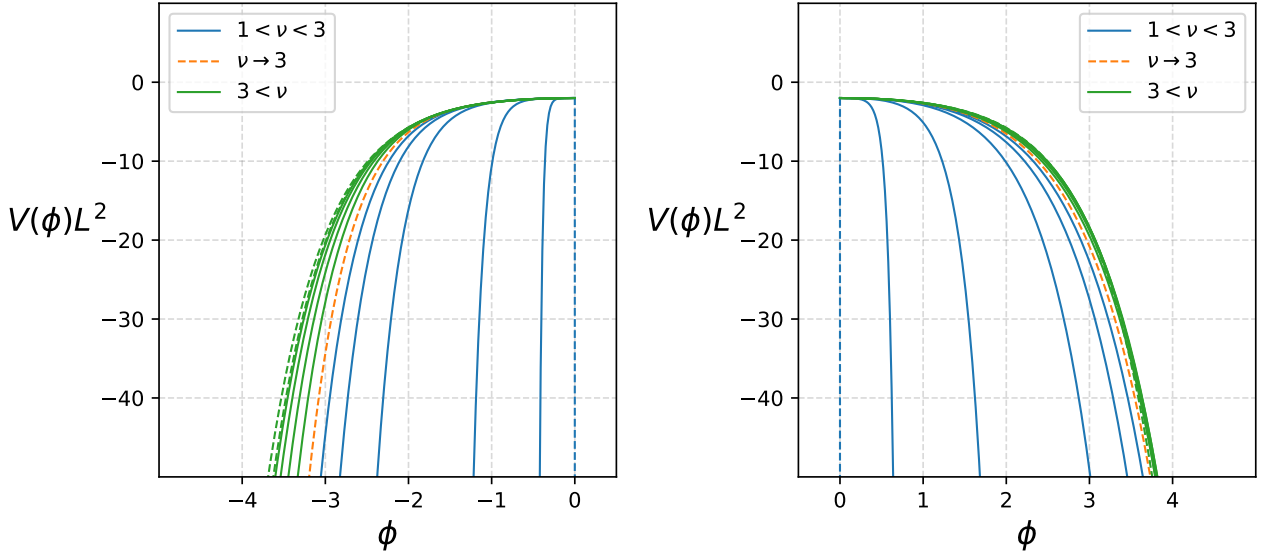

    \centering
    \includegraphics[width=0.5\linewidth]{g1.jpg}\includegraphics[width=0.5\linewidth]{g2.jpg}
    \caption{
The existence of the horizon is ensured by $\alpha L^{2}>0$ for both branches, as demonstrated in the subsection \ref{soluADS}. In the current figures, we have a concave scalar potential for each branch with $alpha L^2=9$. Then the horizon's existence and concavity of the scalar potential agree if  $\alpha L^{2}>0$}
    \label{fig:1}
\end{figure}
\subsection*{The no-hair limit}
The non-hair limit is well defined by $\nu=1$, the scalar field is $\phi=0$, the scalar potential is $V=-2/L^{2}$, and the metric functions are
    \begin{equation}
        \Omega(x)=\frac{1}{\eta^{2}(x-1)^{2}}=r^{2}, \quad f(x)=\frac{1}{L^{2}}-\frac{\alpha(x-1)^{2}}{8} \label{rcoordinate}
    \end{equation}
    in radial coordinates $\Omega(x)=r^{2}$, we have $x=1\pm 1/\eta r$
    \begin{equation}
    -g_{tt}=\Omega(x)f(x)\vert_{x=1\pm 1/\eta r}=-\mu+\frac{r^{2}}{L^{2}}, \qquad
        \mu=-\frac{\alpha}{8\eta^{2}} \label{gttcoordinate}
    \end{equation}
%
\subsection{Thermal superpotential}
\label{TSup}
%
Gursoy and colleagues proposed the existence of a \textit{thermal superpotential} $\mathcal{W}(\phi)$ TS \cite{Gursoy:2008za} in order to transform the second order equation for the scalar field $\phi(x)$ into two first-order equations. One of the main applications of the TS is to regularize the on-shell action \cite{Anabalon:2020qux} and, therefore, the Bronw-York quasilocal stress tensor. From Einstein's equations given in (\ref{einsxx}) we have
\begin{equation}
\label{Wphi}
(\phi^{\prime})^{2}=\frac{3\Omega^{\prime 2}-2\Omega\Omega^{\prime\prime}}{2\Omega^{2}} ~~\Longleftrightarrow ~~    -\frac{\Om^{\prime}}{\eta\Om^{3/2}}=\mathcal{W}, \quad    \frac{\phi^\prime}{\eta\sqrt{\Om}}=\frac{d\mathcal{W}}{d\phi}
\end{equation}
where $\mathcal{W}(\phi)$ is a TS.
And the scalar potential $V(\phi)$
in terms of TS is
\begin{equation}
    V=\frac{1}{2}\[\(\frac{d\mathcal{W}}{d\phi}\)^2-\mathcal{W}^2\]f+\frac{\mathcal{W}}{\eta\sqrt{\Om}}f^{\prime}-\frac{f^{\prime\prime}}{\eta^2\Om}
\end{equation}
There is no event horizon if:\footnote{In the appendix \ref{apendiceA} we have another example in which there is no horizon but $\alpha\neq 0$ and $f(x)$ is non-trivial. But the problem is that space-time has a naked singularity} $\alpha=0$, then $f(x)=1/L^2$, $f^{\prime}(x)=0$ and $f^{\prime\prime}(x)=0$; This yields $V(\phi)$, which is commonly found in supergravity theories \cite{Anabalon:2020qux, Anabalon:2020pez,Anabalon:2017yhv}  
\begin{equation}
    V=\frac{1}{2L^{2}}\[\(\frac{d\mathcal{W}}{d\phi}\)^2-\mathcal{W}^2\]
\end{equation}
From (\ref{Wphi}) we can integrate 
the thermal superpotential for positive $\mathcal{W}_{+}(\phi)$ and negative $\mathcal{W}_{-}(\phi)$ branch   respectively 
\begin{equation}
\label{W1}
    \mathcal{W}_{+}(\phi)=\frac{1}{\nu}\[(\nu+1)e^{\phi\ell(\frac{\nu-1}{2})}+(\nu-1)e^{-\phi\ell(\frac{\nu+1}{2})}\]
\end{equation}
\begin{equation}
\label{W2}
    \mathcal{W}_{-}(\phi)=-\frac{1}{\nu}\[(\nu+1)e^{\phi\ell(\frac{\nu-1}{2})}+(\nu-1)e^{-\phi\ell(\frac{\nu+1}{2})}\]
\end{equation}
The asymptotic expansions at boundary $\phi=0$ are  
\begin{equation}
\label{sup1}
    \mathcal{W}_{+}(\phi)=2+\frac{\phi^{2}}{2}+O(\phi^3)
\end{equation}
\begin{equation}
\label{sup2}
    \mathcal{W}_{-}(\phi)=-2-\frac{\phi^{2}}{2}+O(\phi^3)
\end{equation}
\subsection{Asymptotic boundary conditions}
\label{ABC}
%
We consider $\Omega(x)=r^{2}+\mathcal{O}(r^{-2})$ to get the asymptotic expansion of the metric in radial coordinates, and from that we get the following coordinate change: for the positive branch, the sings are $(+-+)$, and for the negative, $(-++)$: 

\begin{equation}
    x=1\pm \frac{1}{\eta r}\mp\frac{\nu^{2}-1}{24\eta^{3}r^{3}}+\frac{\nu^{2}-1}{24\eta^{4}r^{4}}+\mathcal{O}(r^{-5})
    \label{equ3.18}
\end{equation}
The fall-off for the scalar field $\phi(x)$ in radial coordinates for negative and positive branches, respectively, shows that we have Neumann boundary conditions
\begin{equation}
    \phi(r)=\frac{\sigma_{1}}{r^{\Delta_{+}}}+\frac{\sigma_{2}\ln{(r)}}{r^{\Delta_{-}}}+\mathcal{O}(r^{-2})=\mp\frac{1/\eta\ell}{r}+\frac{0}{r}+\mathcal{O}(r^{-2})
    \label{fall}
\end{equation}
\begin{equation}
    \phi<0 ~\Rightarrow~ \sigma_{1}=-\frac{1}{\eta\ell}, \qquad \sigma_{2}=0
\label{nepo1}
\end{equation}
\begin{equation}
\phi>0 ~\Rightarrow~ \sigma_{1}=\frac{1}{\eta\ell},
\qquad \sigma_{2}=0
\label{nepo2}
\end{equation}
In $D=2+1$ space-time dimensions, the Breitenlohner-Freedman lower bound and unitary upper bound \cite{Breitenlohner:1982bm, Henneaux:2006hk} is $-1\leq m^2 L^{2}<0$. The conformal mass of our solution is $m^{2}L^{2}=-1$, which saturate the BF bound; consequently, the relevant fall-off for the scalar field fix $\Delta_{+}=\Delta_{-}=1$ (\ref{fall}). Relevant fall-off for metric functions derived from asymptotic symmetries $\mathcal{L}_{\xi}{g_{\mu\nu}}=\mathcal{O}(h_{\mu\nu})$ are deviations from $AdS_{3}$ metric. It was proved in \cite{Brown:1986nw,Henneaux:2006hk,Anabalon:2015xvl} that the relevant deviations $h_{\mu\nu}$ in $D=2+1$ are
\begin{equation}
      h_{tt}~=~\mathcal{O}(1), \qquad
    h_{rr}~=~\mathcal{O}(r^{-4})
\end{equation}
The first relevant deviation from $AdS_3$ is $h_{tt}=\mathcal{O}(1)$, indicating that it must be a constant. The second relevant deviation, $h_{rr}=\mathcal{O}(r^{-4})$, implies that the back-reaction of the scalar field gives us the relevant term $\frac{L^{4}}{r^{4}}\qty(\ldots)$. 
Relevant asymptotic metric functions for negative $\phi<0$ and positive $\phi>0$ branches are as follows:

%
\begin{equation}
    -g_{tt}=\Omega(x)f(x)\vert_{x=x(r)}=-\mu+\frac{r^{2}}{L^{2}}+\mathcal{O}(r^{-2}), \qquad
    \mu={\frac {\alpha\, }{{8\eta}^{2}}}
    \label{neg1}
\end{equation}
\begin{equation}
    g_{rr}=\frac{L^{2}}{r^{2}}-\frac{L^{4}}{r^{4}}\qty(-\mu+\frac{\sigma_{1}^{2}}{2L^{2}})
    +\mathcal{O}(r^{-5})
    \label{neg2}
\end{equation}
\begin{equation}
g_{\varphi\varphi}=r^{2}+\mathcal{O}(r^{-5})
\label{neg3}
\end{equation}
$\sigma_1$ was defined in (\ref{nepo1}), (\ref{nepo2}), and it is interesting to note that in $D=3+1$, the asymptotic metric functions for each branch are distinct, whereas in $D=2+1$, this is not the case \cite{Anabalon:2012ta}
%
\section{On-shell finite action}
\label{sec:4}
%

We obtain the finite on-shell action and, consequently, the free energy of the hairy black hole in the current section. It is necessary to supplement the action of the theory (\ref{action1}) with gravitational counterterms $I_{ct}$ and scalar counterterms $I_\phi$ \cite{Aharony:2015afa,Anabalon:2015xvl}. The problem is that the $x$-coordinate is abstract, so we will use radial coordinates; the relevant fall-off was discussed in the previous section \ref{ABC}


%
\begin{equation}
\label{radialG}
    ds^{2}=-\qty[-\mu+\frac{r^{2}}{L^{2}}+\mathcal{O}(r^{-2})]dt^{2}+\qty[\frac{L^{2}}{r^{2}}-\frac{L^{4}}{r^{4}}\qty(-\mu+\frac{\sigma_{1}^{2}}{2L^{2}})
    +\mathcal{O}(r^{-5})]dr^{2}+\qty[r^{2}+\mathcal{O}(r^{-5})]d\varphi^{2}
\end{equation}
The first two contributions to Euclidean on-shell action (\ref{action1}) are

%

\begin{equation}
\label{coun}
I_{bulk}^E+I_{GH}^E=-\beta T\mathcal{S} +\frac{\beta}{4G_N}\qty(\mu-\frac{r^2}{L^2}-\frac{\sigma_{1}^2}{4L^2})
\end{equation}
The gravitational counterterm can eliminate the blowing up quadratic term $r^{2}/L^{2}$
\begin{equation}
\label{gravcon}
    I_{ct}^E=\frac{1}{8\pi G_N}\int_{\partial\mathcal{M}}d^2x\sqrt{h^E}\(\frac{1}{L}\)=\frac{\beta}{4G_{N}}\qty(-\frac{1}{2}\mu+\frac{r^2}{L^2})
\end{equation}
The initial tree contributions result in a finite on-shell action
\begin{equation}
\label{finite3}
I_{bulk}^E+I_{GH}^E+I_{ct}^E=-\beta T\mathcal{S} +\frac{\beta}{4G_N}\qty(\frac{\mu}{2}-\frac{\sigma_{1}^2}{4L^{2}})
\end{equation}
%
%
where the temperature $T=\beta^{-1}$ and entropy $\mathcal{S}$ are
\begin{equation}
    T=\frac{1}{4\pi}\[\frac{(-g_{tt})^\prime}{\sqrt{-g_{tt}g_{rr}}}\]_{r_h},\qquad \mathcal{S}=\(\frac{\sqrt{g_{\varphi\varphi}}\pi}{2G_N}\)_{r_h}
    \label{4.5}
\end{equation}
We could proclaim that the finite on-shell action (reffinite3) is the free energy $F=\beta^{-1}(I_{bulk}^E+I_{GH}^E+I_{ct}^E)$, but the scalar field has a contribution that relates to the well-posed action principle; this boundary term is known as the scalar field contribution
\begin{equation}
    \delta I=-\frac{1}{16\pi G_N}\int{d^{2}x}\sqrt{-h}~n^{\alpha}\partial_{\alpha}\phi~\delta\phi+\delta I_{\phi}
\end{equation}
In order to keep $\delta I_{\phi}=0 $ we choice
\begin{equation}
\delta I_{\phi}=\frac{1}{16\pi G_N}\int{d^{2}x}\sqrt{-h}~n^{\alpha}\partial_{\alpha}\phi~\delta\phi
\end{equation}
In the Euclidean section, the contribution at the boundary $r=\infty$ is finite 
\begin{equation}
    \delta I_{\phi}^{E}=-\frac{1}{16\pi G_N}\int{d^{2}x}\sqrt{h^{E}}~n^{\alpha}\partial_{\alpha}\phi~\delta\phi =\frac{\beta}{8G_N}\(\frac{\sigma_1}{L^2}\)\delta\sigma_1+\mathcal{O}(r^{-2})
\end{equation}
Integrating the extra boundary term yields $I_{\phi}^{E}=\frac{\beta}{4G_N}\qty(\sigma_{1}^{2}/4L^{2})$ which we can add to (\ref{finite3}). However, we'd like to find $I_{\phi}$ precisely instead of $\delta I_{\phi}$; therefore, we propose the following counterterm for the scalar field
\begin{equation}
\label{scalarI}
    I_{\phi}^E=\frac{1}{8\pi G_N}\int_{\partial\mathcal{M}}d^2x\sqrt{h^{E}}\frac{\phi^2}{4L}
    =\frac{\beta}{4G_N}\qty(\frac{\sigma_{1}^{2}}{4L^{2}})+\mathcal{O}(r^{-2})
\end{equation}
When we add this counterterm $I^{E}_{\phi}$ to (\ref{finite3}), we obtain
\begin{equation}
\label{F2}
    I^E=\beta\(M-T\mathcal{S}\), \qquad 
    M=\frac{\mu}{8G_N}
\end{equation}
and the mass of the hairy black hole is $M=\frac{\mu}{8G_N}$. We propose $\mathcal{W}(\phi)$ as a counterterm, which is equivalent to gravitational (\ref{gravcon}) and scalar (\ref{scalarI}) counterterms \cite{Anabalon:2020qux}
\begin{equation}
	I^{E}_{\mathcal{W}(\phi)}=\frac{1}{8\pi G_N}\,\int_{\partial M} d^{2}x\,\sqrt{h}\,\left[
	\frac{\mathcal{W}_+(\phi)}{2L}\right]\;=\frac{\beta}{4G_N}\left(-\frac{1}{2}\mu+\frac{r^2}{L^2}+\frac{\sigma_1^2}{4L^2} \right)
	\label{onshellsuper}
\end{equation}
\begin{equation}
	I^{E}_{\mathcal{W}(\phi)}=-\frac{1}{8\pi G_N}\,\int_{\partial M} d^{2}x\,\sqrt{h}\,\left[
	\frac{\mathcal{W}_-(\phi)}{2L}\right]\;=\frac{\beta}{4G_N}\left(-\frac{1}{2}\mu+\frac{r^2}{L^2}+\frac{\sigma_1^2}{4L^2} \right)
	\label{onshellsuper1}
\end{equation}

\begin{equation}
I^{\textsc{e}}_\text{bulk}+I^{\textsc{e}}_{\textsc{gh}}+I^{E}_{\mathcal{W}(\phi)}=\beta\(M-T\mathcal{S}\)
\label{4.13}
\end{equation}
In the following section, we construct the regularized holographic stress tensor with our counterterm (in the Lorentzian section)
%
\section{Holographic stress tensor}
\label{sec:5}
The holographic stress tensor or Brown-York stress tensor
was defined in \cite{Brown:1992br}. It is built upon the time-like foliation $x=constant$. We prefer working with radial coordinates, taking into account the relevant falloff (\ref{radialG})
\begin{equation}
\tau_{ab}=\frac{2}{\sqrt{-h}}\frac{\delta I}{\delta h^{ab}}
\end{equation}
To obtain a finite holographic stress tensor $\tau_{ab}$, a regularized action in the Lorentzian section is required
\begin{equation}
    I[g_{\mu\nu},\phi]=I_{bulk}+\frac{1}{8\pi G_{N}}%
    \int_{\partial\mathcal{M}}{d^{2}xK\sqrt{-h}}-\frac{1}{8\pi G_{N}}\int_{\partial\mathcal{M}}d^2x\sqrt{-h}\frac{1}{L}-\frac{1}{8\pi G_{N}}\int_{\partial\mathcal{M}}d^2x\sqrt{-h}\frac{\phi^2}{4L}
    \label{action2}
\end{equation}
Here $h_{ab}$ represents the time-like induced metric, so we have
\begin{equation}
\label{tau1}
    \tau_{ab}=-\frac{1}{8\pi G_{N}}\(K_{ab}-h_{ab}K+\frac{h_{ab}}{L}\)-\frac{h_{ab}}{32\pi G_{N}}\dfrac{\phi^{2}}{L}
\end{equation}
At the boundary $r=\infty$, the components that are finite are
\begin{align}
\label{stressBK}
 &\tau_{tt}=\frac{1}{8\pi G_{N}}\(\frac{\mu}{2L}\)+\mathcal{O}(r^{-2})\\ \nonumber
&\tau_{\varphi\varphi}=\frac{1}{8\pi G_{N}}\(\frac{\mu L}{2}\) + \mathcal{O}(r^{-2})\\ \nonumber
\end{align}
Locally, the boundary metric (\ref{radialG}) can be written in ADM-like: $ds^{2}=\frac{r^{2}}{L^{2}}\qty(-dt^{2}+L^{2}d\varphi^{2})$. The isometry for this boundary metric is generated by the killing vector $\xi^{b}=\delta^{b}_{t}$, and the quantity associated with the temporal isometry that is conserved is the energy\footnote{This confirms that our previous result (\ref{F2}) had been accurate}
\begin{equation} E=\oint_{\Sigma}dy\sqrt{\sigma}u^a\tau_{ab}\xi^b=\frac{\sigma \sqrt{S}}{\sqrt{N}}\tau_{tt}=\frac{ \mu}{8G_N}+\mathcal{O}(r^{-2})
\end{equation}
From the boundary metric $ds^{2}=\frac{r^{2}}{L^{2}}\qty(-dt^{2}+L^{2}d\varphi^{2})$ up to the conformal factor $r^{2}/L^{2}$, we can get the background metric $\gamma_{ab}dx^{a}dx^{b}=-dt^{2}+L^{2}d\varphi^{2}$ in which the dual quantum field theory is defined.
An important observable in the dual field theory is the stress tensor $\expval{\tau_{ab}}$, which in general
is related to the Brown-York stress tensor $\tau_{ab}$ up to conformal factor \cite{Lola:1998cr}, but in $D=2+1$ is not necessary, because (\ref{stressBK}) is finite at boundary $r=\infty$, then
\begin{equation}
\expval{\tau_{ab}}=\lim_{r\rightarrow\infty}\tau_{ab} ~\Rightarrow ~ \expval{\tau_{ab}}=  \tau_{ab}
\end{equation}
the boundary metric $\gamma_{ab}$
is non-dynamical, and we can write  $\expval{\tau_{ab}}$ can be expressed as a perfect fluid
\begin{equation}
    \expval{\tau_{ab}}=(\rho+p)u_{a}u_{b}+p\gamma_{ab}
\end{equation}
where the density and pressure are
\begin{equation}
    \rho=\frac{1}{16\pi G_N}\(\frac{\mu}{ L}\), \qquad p=\frac{1}{16 \pi G_N}\(\frac{\mu}{ L}\)
\end{equation}
Conforming to expectations, our solution preserves conformal symmetry and the dual stress tensor's trace is null
\begin{equation}
\gamma^{ab}\expval{\tau_{ab}}=-\rho +p=0
\end{equation}
Last but not least, we would like to show an alternative holographic stress tensor to (\ref{tau1}). If we consider the thermal superpotential instead of gravitational and scalar counterterms, then our regularized action is
\begin{equation}
    I[g_{\mu\nu},\phi]=I_{bulk}+\frac{1}{8\pi G_{N}}%
    \int_{\partial\mathcal{M}}{d^{2}xK\sqrt{-h}}-\frac{1}{8\pi G_{N}}\int_{\partial\mathcal{M}}d^2x\sqrt{-h}\qty[\pm\frac{\mathcal{W_{\pm}}(\phi)}{2L}]
    \label{action3}
\end{equation}
Consequently, our alterative holographic stress tensor is
\begin{equation}
\label{tau2}
    \tau_{ab}=-\frac{1}{8\pi G_{N}}\(K_{ab}-h_{ab}K\)\mp\frac{h_{ab}}{8\pi G_{N}}\dfrac{\mathcal{W}_{\pm}(\phi)}{2L}
\end{equation}
Moreover, the components $\tau_{ab}$ at the boundary are the same to those obtained previously (\ref{stressBK})
%
\section{Thermodynamics}
\label{sec:6}
%
In this section, we verify the first law of thermodynamic quantities and construct the $T~vs~S$ and $F~vs~T$ phase diagrams. Since the mass expression is identical for both branches, $\eta(M)$ is replaced for $M(\eta)$
\begin{equation}
M=\frac{\alpha}{64\eta^2G_{N}} ~~\Rightarrow~~
\eta=\frac{1}{8}\sqrt{\frac{\alpha}{G_NM}}
\end{equation}
\subsubsection*{Negative branch:}
$\phi <0$ or $0<x\leq 1$. The temperature,
mass, and entropy are
\begin{equation}
    T=\[\frac{f(x)^{\prime}}{4\pi\eta}\]_{x_{h}}=\frac{\alpha}{16\eta\nu\pi}\(x_h^{\frac{1-\nu}{2}}-x_h^{\frac{1+\nu}{2}}\)
    \label{equ6.1}
\end{equation}
\begin{equation}
 \mathcal{S}=\frac{\nu\pi x_{h}^{\frac{\nu-1}{2}}}{2\eta G_{N} (1-x^{\nu}_{h})}, \qquad~~ M=\frac{\mu}{8 G_{N}}=\frac{\alpha}{64\eta^2 G_{N}}
 \label{equ6.2}
\end{equation}
All of these thermodynamic quantities obey the first law: $dM=Td\mathcal{S}$, and the free energy $F=\beta^{-1}I^{E}$ has been derived from the on-shell finite action (\ref{F2})
\begin{equation}
F=M-T\mathcal{S}=-\frac{\alpha}{64\eta^2G_{N}}
\label{equ6.3}
\end{equation}
and it is simple to verify $T\mathcal{S}=2M$ hence $F=-M$.
The heat capacity is $C=dM/dT$ 
\begin{equation}
 C=\frac{\nu\pi}{2\eta G_N\(x_h^{\frac{1-\nu}{2}}-x_h^{\frac{1+\nu}{2}}\)}=
 \mathcal{S}
\end{equation}
In conclusion, the mass-parameterized thermodynamic quantities in the negative branch are
\begin{equation}
T=\frac{\sqrt{\alpha}~\mathcal{X}(x_{h},\nu)}{2\pi\nu}\sqrt{MG_{N}}, \qquad \mathcal{S}=\frac{4\pi\nu \mathcal{X}(x_{h},\nu)^{-1}}{G_{N}\sqrt{\alpha}}\sqrt{M G_{N}}, \qquad F=-M
\end{equation}
where $\mathcal{X}(x_{h},\nu)$ can be fixed from horizon equation $f(x_{h},\nu)=0$
\begin{equation}
\mathcal{X}(x_{h},\nu)=x_h^{\frac{1-\nu}{2}}-x_h^{\frac{1+\nu}{2}}
\end{equation}
\subsection*{Positive branch:}
\begin{figure}[h!]
\centering
    \includegraphics[width=0.45\linewidth]{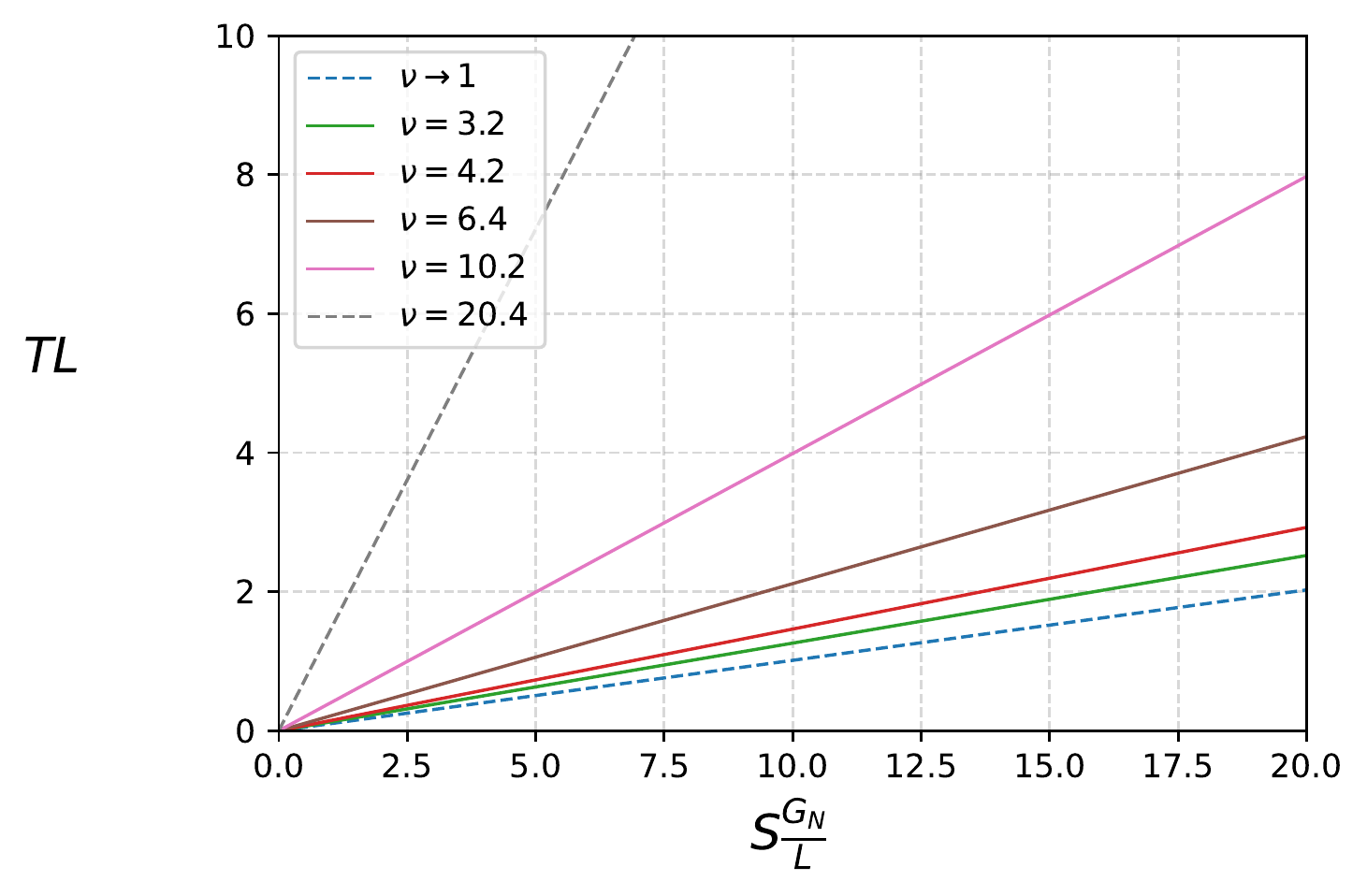}
    \includegraphics[width=0.45\linewidth]{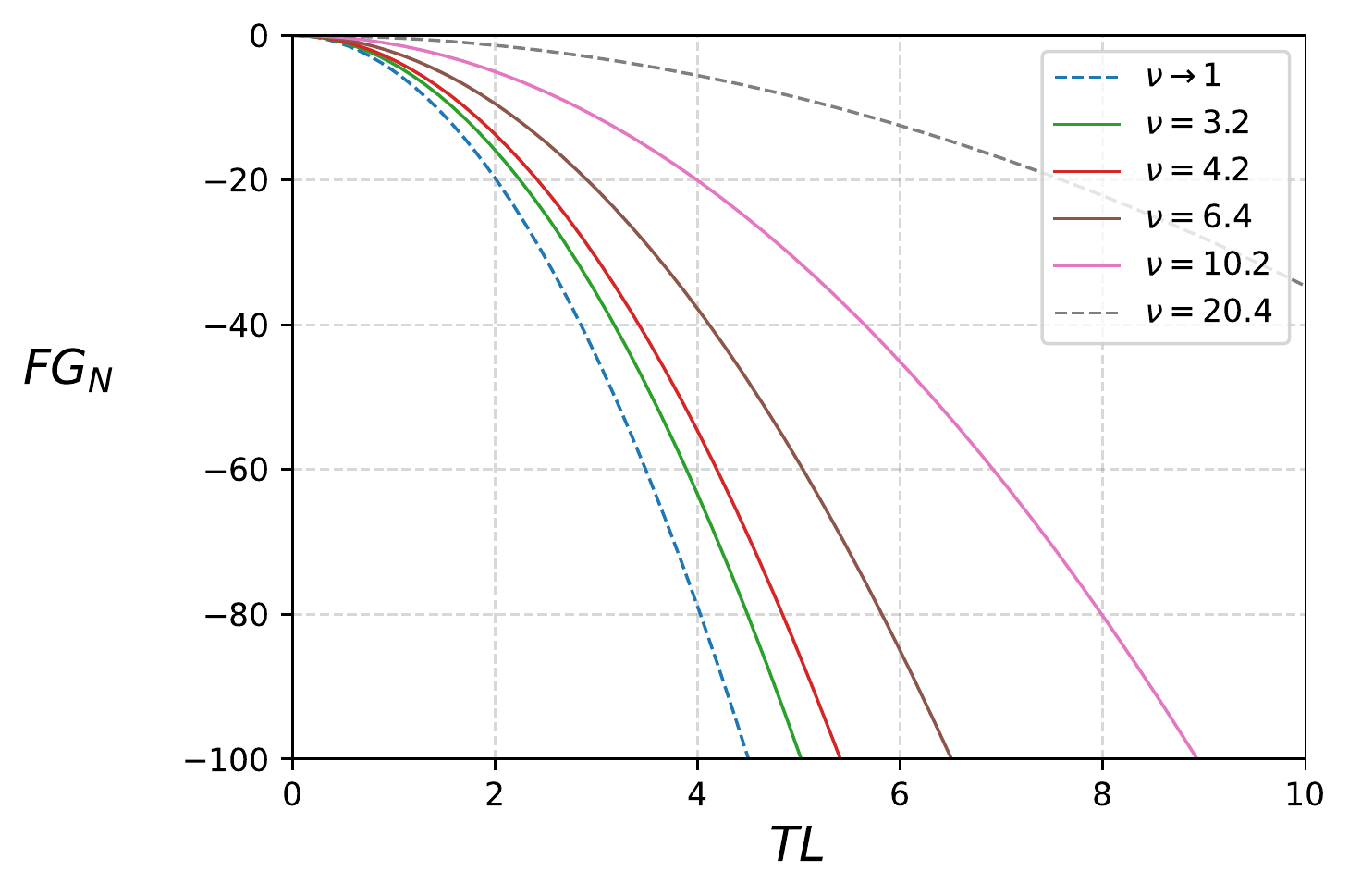}
    \caption{In this pair of figures, we examine the behavior of the following phase diagrams ($T~vs~S$ and $F~vs~T$) for various values of the hairy parameter $\nu$; $\alpha=9/L^2$ is fixed. These illustrations are applicable to both the positive and negative branches. The blue dashed line represents the behavior of these thermodynamic quantities when the value of the hairy parameter is $\nu=1$, i.e. $Schwarzschild-AdS_{2+1}$.
    }
    \label{fig:3}
\end{figure}

$\phi>$ or $1\leq x<\infty$. The sign of temperature and entropy will change, but all other thermodynamic quantities $M, F, C$ remain unchanged
\begin{equation}
T=-\frac{\sqrt{\alpha}~\mathcal{X}(x_{h},\nu)}{2\pi\nu}\sqrt{MG_{N}}, \qquad \mathcal{S}=-\frac{4\pi\nu \mathcal{X}(x_{h},\nu)^{-1}}{G_{N}\sqrt{\alpha}}\sqrt{M G_{N}}, \qquad F=-M
\label{equ6.5}
\end{equation}
Finally, we would like to mention the 
non-hair limit. 
If $\nu=1$ is fixed in the horizon equation $f(x_{h},\alpha)=0$ , then we obtain $x_{h}$ for negative and positive branches, respectively
\begin{equation}
x_{h}=1-\frac{2\sqrt{2}}{L\sqrt{\alpha}}, \qquad
x_{h}=1+\frac{2\sqrt{2}}{L\sqrt{\alpha}}
\end{equation}
By replacing $x_{h}$ in the hairy thermodynamic quantities for both branches, Schwarzschild results are obtained $AdS_{2+1}$, see appendix \ref{apendiceB} 
\begin{equation}
T=\frac{\sqrt{2}}{\pi L}\sqrt{M_{0}G_{N}}, \qquad
\mathcal{S}=\frac{\pi L\sqrt{2}}{G_{N}}\sqrt{M_{0}G_{N}}, \qquad
F=-M_{0}
\end{equation}
To work only with dimensionless quantities, the following rescaling is applied
\begin{equation}
    T \rightarrow TL \qquad M\rightarrow MG_{N} \qquad F\rightarrow FG_{N} \qquad \mathcal{S}\rightarrow \frac{\mathcal{S}G_{N}}{L}
\end{equation}
In thermodynamics, phase diagrams of the thermodynamic quantities $(T, \mathcal{S}, F)$ are a very useful tool for illustrating the system's behavior. There is no essential difference between the negative and positive branch diagrams; see figure \ref{fig:3}. 

Unstable states are characterized by a rapid increase in temperature without a maximum peak; the $T~vs~S$ diagram in figure \ref{fig:3} illustrates that when the hair is more extensive $\nu>>1$, the slope is higher; therefore, we can conclude that hairy black holes are more unstable than hairless black holes under thermal fluctuations. The figure on the left of \ref{fig:3} depicts $F~vs~T$; it demonstrates that the free energy of hairy black holes $F_{hairy}(\nu)$ is less than that of hairless black holes $F_{Sch-AdS}(\nu=1)$; therefore, hairless black holes are preferable to hairy black holes

\newpage
\section{Discussion}
\label{sec:7}
%
The hairy solution was constructed as follows: the input is (\ref{om1}), from which the scalar field $\phi(x)$, the metric function $f(x)$, and the scalar potential (\ref{pot}) can be integrated. Consequently, the scalar potential was constructed on-shell. The election of $\Omega(x)$ was first presented in the context of an integrable system with two killing vectors by \cite{Anabalon:2012dw}.
The constant of the theory, $\alpha$, plays two crucial roles: first, it ensures the concavity of the scalar potential, see figure \ref{fig:1}, and second, it controls the existence of the horizon $f(x_h,\alpha)=0$. In fact, in (\ref{soluADS}), we verify that $\alpha L^{2}>0$, indicating that $\alpha$ controls the back-reaction of the scalar field; see (\ref{pot}). Concerning the existence of the event horizon, in \cite{Anabalon:2013sra}, they demonstrated that an asymptotically flat solution can be obtained by properly setting $\Lambda=0$; if we apply this procedure to our hairy solution, we obtain a naked singularity \ref{apendiceA}, suggesting that the back reaction of the scalar field is insufficient to curve space-time and generate an event horizon.

Using the well-posed variational principle, we construct the scalar field's counterterm. Surprisingly, the $\mathcal{W}_{\pm}(\phi)$ thermal superpotential can also be used as a counterterm. There is no detailed description of why the thermal superpotential features as a counterterm of the scalar field at this time,
see another example in \cite{Anabalon:2019tcy}.
The asymptotic expansion of the $I_{\mathcal{W}(\phi)}$ reveals the equivalence of the counterterms gravitational $I_{ct}$ and scalar $I_{\phi}$ with the thermal superpotential proposal $I_{\mathcal{W}(\phi)}$
\begin{equation}
I_{ct}+I_{\phi}=-\frac{1}{8\pi G_{N}}\int_{\partial\mathcal{M}}d^{2}x\sqrt{-h}\qty(\frac{1}{L}+\frac{\phi^{2}}{4L})
\end{equation}
\begin{equation}
I_{\mathcal{W(\phi)}}=-\frac{1}{8\pi G_{N}}\int_{\partial\mathcal{M}}d^{2}x\sqrt{-h}~\frac{\mathcal{W}_{+}(\phi)}{2L}=-\frac{1}{8\pi G_{N}}\int_{\partial\mathcal{M}}d^{2}x\sqrt{-h}~\qty(\frac{1}{L}+\frac{\phi^{2}}{4L}+\mathcal{O}(\phi^{3}))
\end{equation}
It would be worthwhile to investigate the relevant fall-off terms of $\mathcal{W}_{\pm}(\phi)$ that regularize the action and Brown-York stress tensor. In accordance with a well-defined variational principle. 

\begin{figure}[H]
    \centering
\begin{tikzpicture}[>=stealth,thick]
    \draw (0,2) -- (5,2) node[right] {$F_{AdS}=0$};
    \draw (0,1) -- (5,1) node[right] {$F_{hairy}$};
    \draw (0,0) -- (5,0) node[right] {$F_{Sch-AdS}$};
    \draw[->] (1.25,2) -- (1.25,1);
    \draw[->] (2.5,2) -- (2.5,0);
    \draw[->] (3.75,1) -- (3.75,0);
\end{tikzpicture}
    \caption{The free energy of the hairy black hole $F_{\text{hairy}}$ and $F_{\text{Sch-AdS}}$ are calculated with respect to ground state $F_{\text{AdS}}$}
    \label{FFF}
\end{figure}
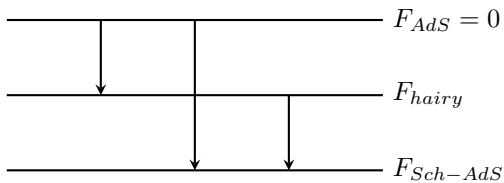

Finally, about thermodynamics, the diagram $F~vs~T$, see figure \ref{fig:3}, shows  that $F_{Sch-AdS}<F_{hairy}$, see figure \ref{FFF}, i.e.  the black hole phase without hair $F_{Sch-AdS}$ is more stable than the hairless phase $F_{hairy}$ leaving open the possibility of second-order phase transitions, see figure \ref{FFF}, which could be explored in a future work
In summary
\begin{equation*}
\left. \begin{array}{ll}
 F_{\text{hairy}}-F_{\text{AdS}} < 0 &\\
 F_{\text{Sch-AdS}}-F_{\text{AdS}} < 0
\end{array} \right\}~~\text{First-Order Phase Transitions}
\end{equation*}

\begin{equation*}
\left. \begin{array}{ll}
 F_{\text{Sch-AdS}}-F_{\text{hairy}} < 0
\end{array} \right\}~~\text{Second-Order Phase Transitions}
\end{equation*}
\section*{\normalsize Acknowledgments}
\vspace{-5pt}

D. Choque would like to thank Raúl Rojas for interesting discussions on hairy black hole solutions. C. Desa, W. Ccuiro and D. Choque, We would like to thank the Universidad Nacional San Antonio Abad del Cusco for the support provided during the development of this work

\clearpage
\begin{appendices}
\renewcommand{\theequation}{\thesection\arabic{equation}}

\section{Integration of the scalar potential}\label{apendiceInt}
\label{apendiceAA}
%
From the second equation of (\ref{einsxx}) and taking $\Omega(x)$ as an input \cite{Anabalon:2013sra,Anabalon:2013qua,Acena:2012mr,Acena:2013jya}:
\begin{equation}
    (\Omega^{1/2}f^{\prime})^{\prime}=0  , \qquad \Om(x)=\frac{\nu^2x^{\nu-1}}{\eta^2(x^\nu-1)^2}
\end{equation}
$\eta$ is an integration constant associated with the black hole's mass. It is simple to integrate $f(x)$, but we must now fix two constants of integration $c_1$ and $c_2$
\begin{equation}
f(x)=\frac{2c_1\eta}{\nu}\(\frac{x^{\frac{3+\nu}{2}}}{3+\nu}-\frac{x^{\frac{3-\nu}{2}}}{3-\nu}\)+c_2
\end{equation}
The metric (\ref{xcoor}) at the boundary $x=1$ must be locally $AdS_2$; therefore, $\Omega(x)=r^{2}$ and $f(x)=1/L^{2}$ are obtained
\begin{equation}
\lim_{x=1}~\Omega(x)\left[  -f(x)dt^{2}+\frac{\eta^{2}dx^{2}}{f(x)}+d\varphi^{2}\right]=-\frac{r^{2}}{L^{2}}dt^{2}+r^{2}d\varphi^{2}~\Rightarrow~
\lim_{x=1}f(x)=\frac{1}{L^{2}}
\end{equation}
It allows us to determine the constant $c_{2}$
\begin{equation}
    \lim_{x\rightarrow 1}f(x)={\frac {{\nu}^{2}{\it c_2}+4\,{\it c_1}\,\eta-9\,{\it c_2}}{{\nu}^{2}-9}}=\frac{1}{L^2} \Rightarrow c_2=\frac{1}{L^2}-\frac{4c_1\eta}{\nu^2-9}
\end{equation}
By substituting $f(x)$ and $\Omega(x)$ in (\ref{einsxx}), the scalar potential $V(x)$ is obtained; this potential is dependent on the constants $\eta$and $c_1$. We proclaim that $\eta$ is related to the mass of the black hole; therefore, $\eta$ should vanish from the potential $V(x)$, as integration constants should not, in principle, appear in the potential of the theory
\begin{multline}
V(x)  =-\frac{3}{4\nu^{2}}\qty(\frac{1}{L^2}-\frac{4c_1\eta}{\nu^2-9})x^{-\frac{\sqrt{2\nu^{2}-2}}{2}}\bl{[}(1+\nu) \bl{(}1+\frac{\nu}{3}\br{)}x^{\frac{\sqrt{2\nu^{2}-2}}{2}}+
(1-\nu)\bl{(}1-\frac{\nu}{3}\br{)}x^{-\frac{\sqrt{2\nu^{2}-2}}{2}}\\
-2(1-\nu^{2})\br{]} +\frac{4\eta c_{1}\cdot x^{\frac{\sqrt{2\nu^{2}-2}}{4}}}{\nu(\nu^2-9)}\[(1-\nu)x^{\frac{\sqrt{2\nu^{2}-2}}{4}}-(1+\nu)x^{-\frac{\sqrt{2\nu^{2}-2}}{4}}\]
\end{multline}
%
%
To eliminate $\eta$ from the scalar potential, we set $c_{1}=-\alpha/4\eta$; $\alpha$ is now a constant of the theory. The constants $c_1$ and $c_2$ are then defined by
\begin{equation}
    c1=-\frac{\alpha}{4\eta}, \qquad 
    c_2=\frac{1}{L^2}+\frac{\alpha}{\nu^2-9}
\end{equation}
%
\section{Asymptotically flat solution}
\label{apendiceA}
%
Here we present the asymptotically flat potential, which is well defined under $L\rightarrow\infty$ in (\ref{pot})
\begin{multline}
    V(\phi)  =-\frac{3\alpha \exp(-\phi\ell)}{4\nu^{2}(\nu^2-9)}\bl{[}(1+\nu) \bl{(}1+\frac{\nu}{3}\br{)}\exp\(\phi\ell\nu\)+
    (1-\nu)\bl{(}1-\frac{\nu}{3}\br{)}\exp\(-\phi\ell\nu\)-2(1-\nu^{2})\br{]}\\
    -\frac{\alpha\exp(\phi\ell /2)}{\nu(\nu^2-9)}\[(1-\nu)\exp\(\frac{\phi\ell\nu}{2}\)-(1+\nu)\exp\(-\frac{\phi\ell\nu}{2}\)\]
\end{multline}
The metric components asymptotically flat can be directly obtained from (\ref{metric11})
at $L\rightarrow\infty$ and the conformal factor $\Omega(x)$ does not change
\begin{equation}
    f(x)=\frac{\alpha}{2\nu}\bl{[}
    \frac{2\nu}{\nu^2-9} -\frac{1}{\(3+\nu\)}x^{\frac{3+\nu}{2}}+\frac{1}{\(3-\nu\)}x^{\frac{3-\nu}{2}}\br{]}
    \label{metricflat}
\end{equation}
The problem is that the metric function $f(x)$ does not have any zeros, since $\frac{f(x)\nu}{\alpha}<0$ is negative definite for $0<x<\infty$ and $1<\nu\leq \infty$. It is evident that the scalar field's back-reaction is insufficient to permit the existence of an event horizon. Therefore, this hairy black hole ($\Lambda=0$) possesses a naked singularity

\newpage

\section{Schwarzschild  $AdS_{2+1}$}
\label{apendiceB}
%
The Einstein-Hilbert action with the Hawkings-Gibbson term, which is an action of surface
\begin{equation}
    I_{bulk}+I_{GH}=\frac{1}{16\pi G_N}\int_{\mathcal{M}}d^{n+1}x\sqrt{-g}\left(R+\frac{n(n-1)}{L^2}\right)+\frac{1}{8\pi G_N}\int_{\partial\mathcal{M}}d^nx\sqrt{-h}K
\end{equation}
We already know the equation of motion of action with cosmological constant without matter
\begin{equation}
    R_{\mu \nu} - \frac{1}{2} g_{\mu \nu} R + \Lambda g_{\mu \nu} = 0
\end{equation}
So, if the cosmological constant is $\Lambda=-\frac{n(n-1)}{2L^2}$.
The thermodynamics of asymptotically planar black holes and $AdS$ in $D=2+1$ or $n=2$ will be investigated.
The solution for the metric is
\begin{equation}
ds^2=-f(r)dt^2+f(r)^{-1}dr^2+r^2d\Sigma_{n-1}^{2}
\end{equation}
\begin{itemize}
    \item \textbf{Asymptotically-$AdS_{2+1}$:}
    The metric function is, where $d\Sigma_{1}^{2}=d\varphi^{2}$ 
    \begin{equation}
    f(r)=-8MG_{N}+\frac{r^{2}}{L^{2}}
    \end{equation}
    The thermodynamic quantities are  
\begin{equation}
M=\frac{r_{h}^{2}}{8G_{N}L^{2}}, \qquad
T=\frac{r_{h}}{2L^{2}\pi}, \qquad
S=\frac{\pi r_{h}}{2G_{N}}, \qquad
F=-\frac{r_{h}^{2}}{8G_{N}L^{2}}, \qquad C=\frac{\pi r_{h}}{2G_{N}}=S
\end{equation}
\end{itemize}
\begin{figure}[ht]
    \includegraphics[width=0.24\linewidth]{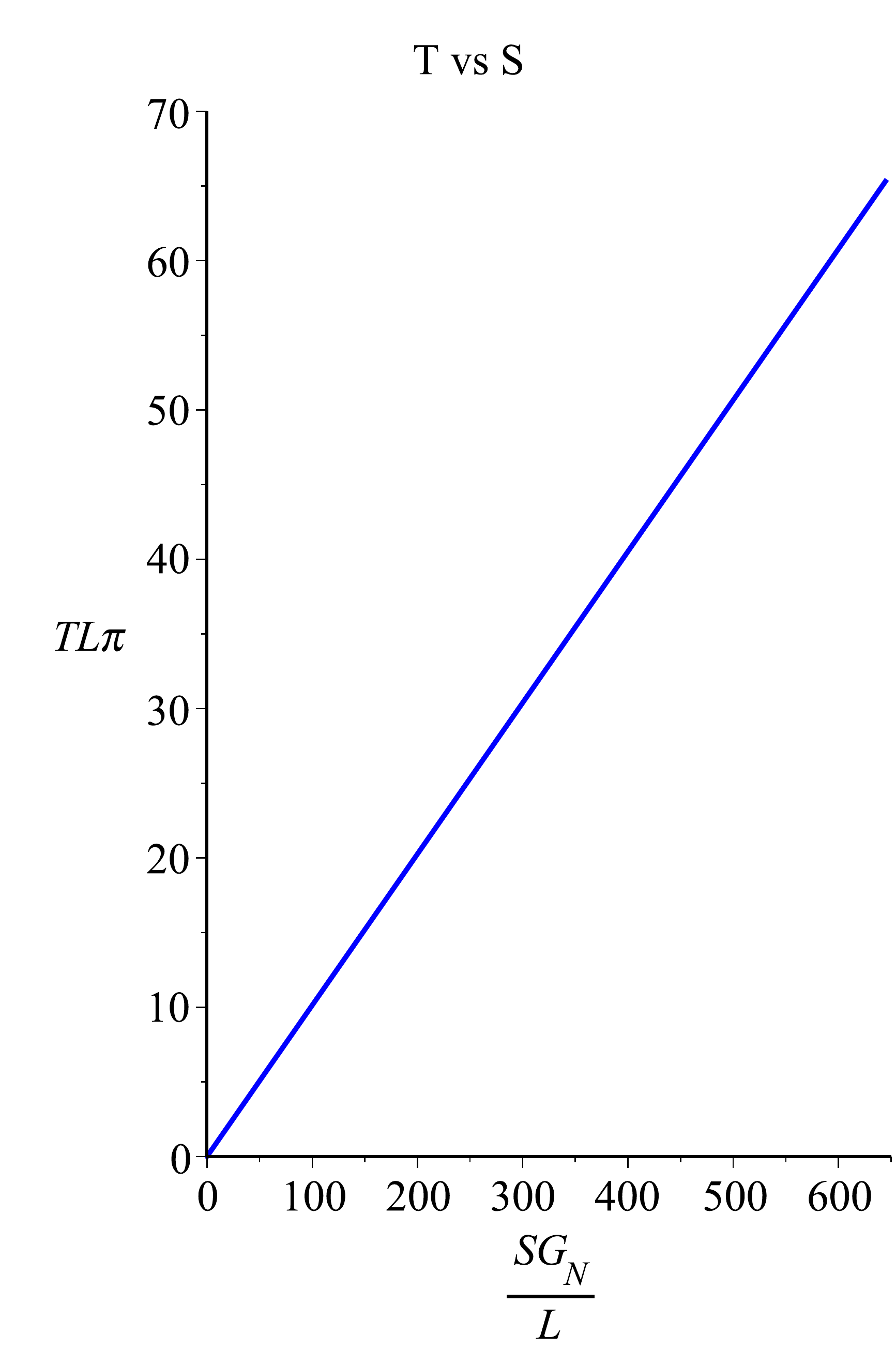}
    \includegraphics[width=0.24\linewidth]{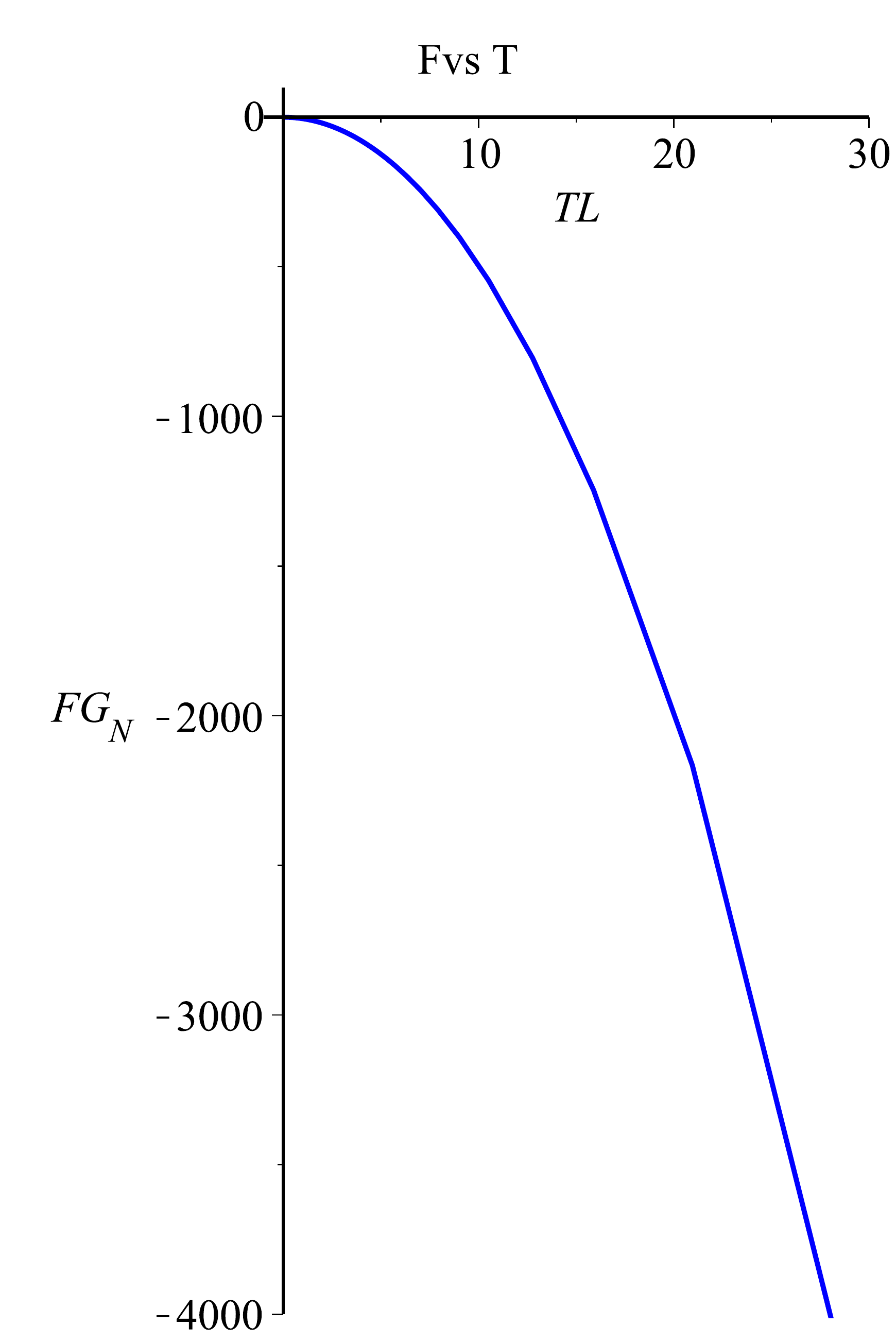}
    \includegraphics[width=0.24\linewidth]{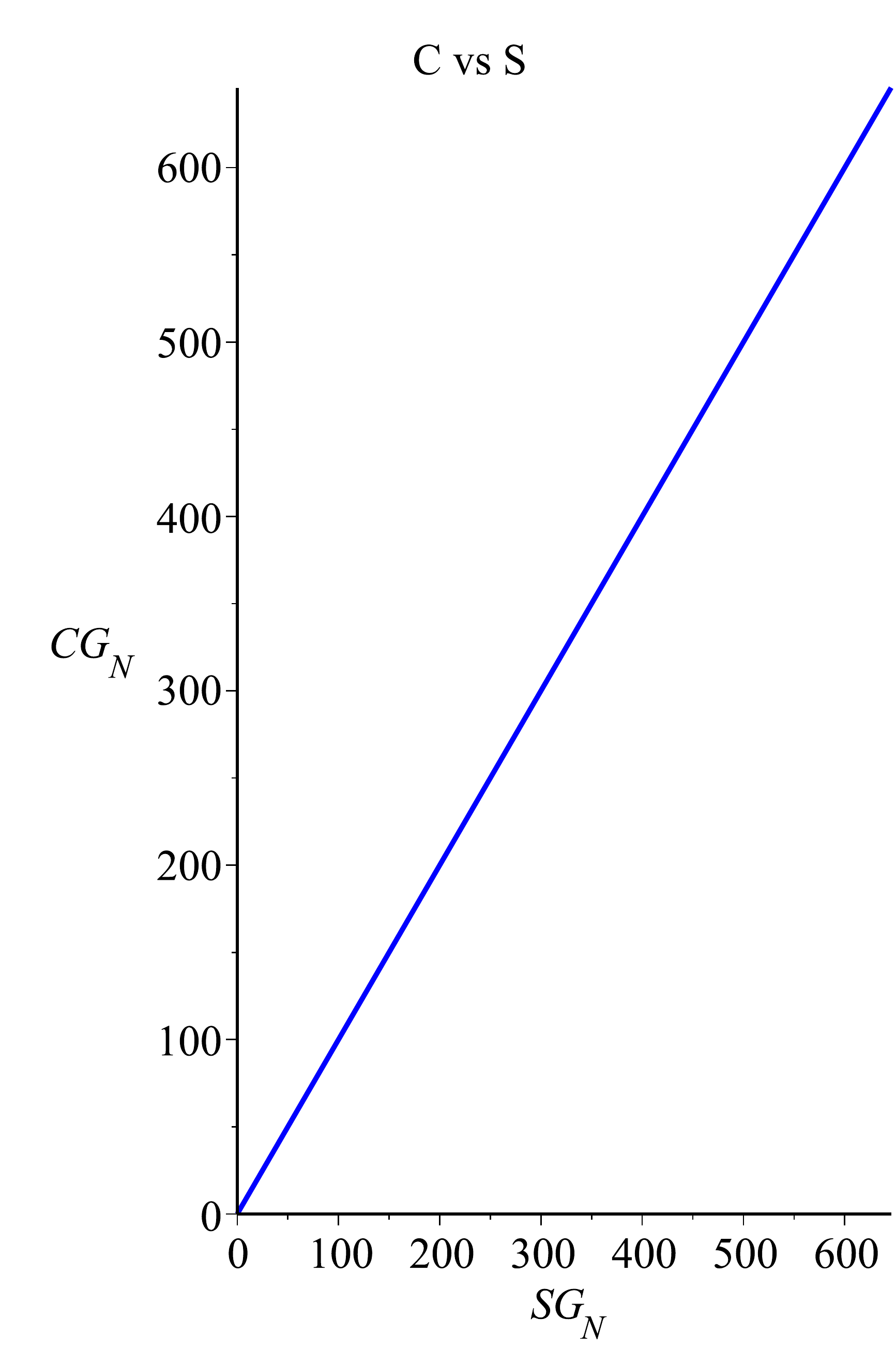}
    \includegraphics[width=0.24\linewidth]{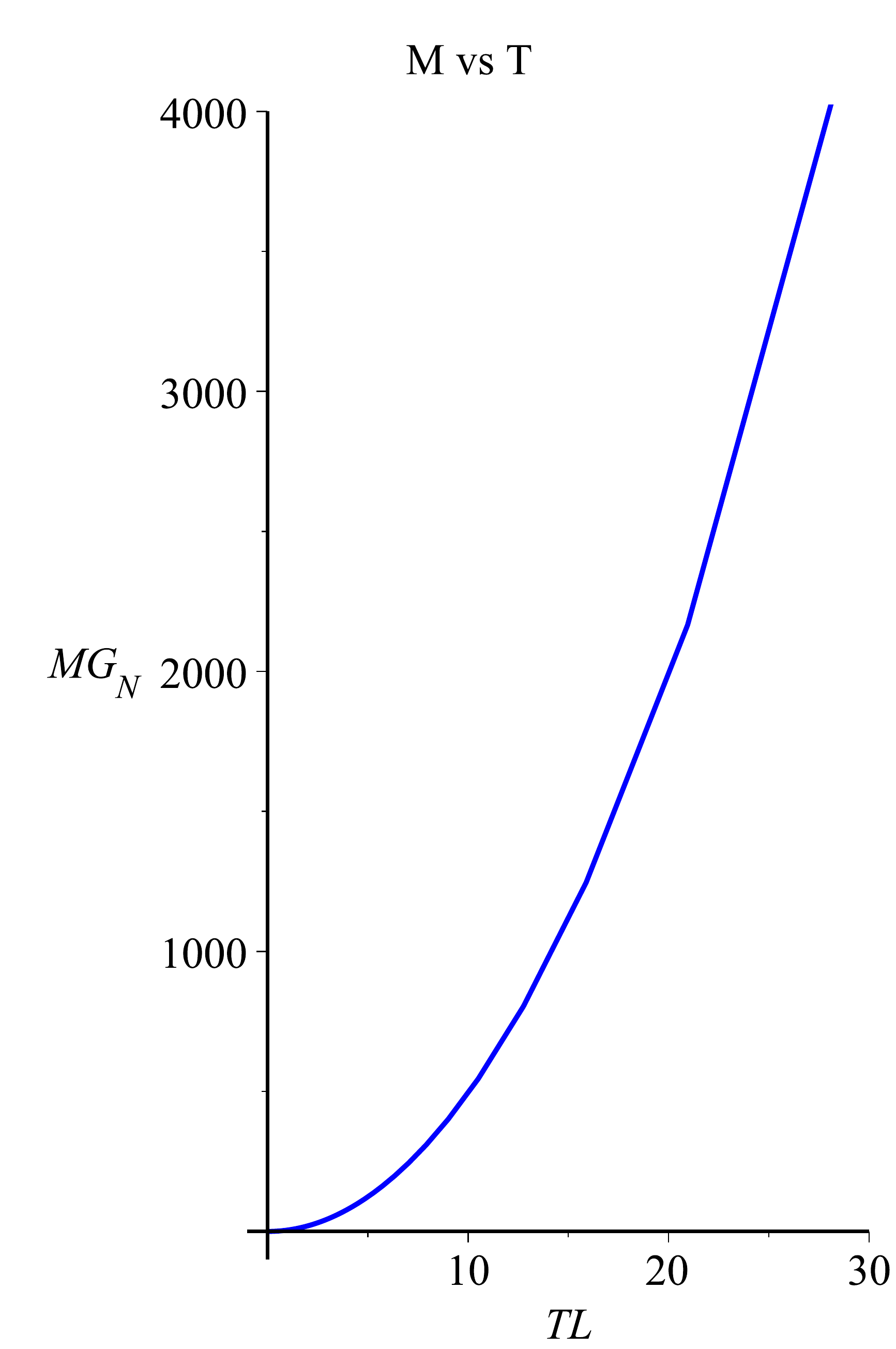}
    \caption{ $AdS_{2+1}$}
    \label{fig:8}
\end{figure}

\end{appendices}

\clearpage
%


\newpage

\hypersetup{linkcolor=blue}
\phantomsection 
\addtocontents{toc}{\protect\addvspace{4.5pt}}
\addcontentsline{toc}{section}{References} 
\bibliographystyle{apsrev4-1}
\bibliography{bibliographyBH} 

\begin{thebibliography}{69}%
\makeatletter
\providecommand \@ifxundefined [1]{%
 \@ifx{#1\undefined}
}%
\providecommand \@ifnum [1]{%
 \ifnum #1\expandafter \@firstoftwo
 \else \expandafter \@secondoftwo
 \fi
}%
\providecommand \@ifx [1]{%
 \ifx #1\expandafter \@firstoftwo
 \else \expandafter \@secondoftwo
 \fi
}%
\providecommand \natexlab [1]{#1}%
\providecommand \enquote  [1]{``#1''}%
\providecommand \bibnamefont  [1]{#1}%
\providecommand \bibfnamefont [1]{#1}%
\providecommand \citenamefont [1]{#1}%
\providecommand \href@noop [0]{\@secondoftwo}%
\providecommand \href [0]{\begingroup \@sanitize@url \@href}%
\providecommand \@href[1]{\@@startlink{#1}\@@href}%
\providecommand \@@href[1]{\endgroup#1\@@endlink}%
\providecommand \@sanitize@url [0]{\catcode `\\12\catcode `\$12\catcode
  `\&12\catcode `\#12\catcode `\^12\catcode `\_12\catcode `\%12\relax}%
\providecommand \@@startlink[1]{}%
\providecommand \@@endlink[0]{}%
\providecommand \url  [0]{\begingroup\@sanitize@url \@url }%
\providecommand \@url [1]{\endgroup\@href {#1}{\urlprefix }}%
\providecommand \urlprefix  [0]{URL }%
\providecommand \Eprint [0]{\href }%
\providecommand \doibase [0]{http://dx.doi.org/}%
\providecommand \selectlanguage [0]{\@gobble}%
\providecommand \bibinfo  [0]{\@secondoftwo}%
\providecommand \bibfield  [0]{\@secondoftwo}%
\providecommand \translation [1]{[#1]}%
\providecommand \BibitemOpen [0]{}%
\providecommand \bibitemStop [0]{}%
\providecommand \bibitemNoStop [0]{.\EOS\space}%
\providecommand \EOS [0]{\spacefactor3000\relax}%
\providecommand \BibitemShut  [1]{\csname bibitem#1\endcsname}%
\let\auto@bib@innerbib\@empty
\bibitem [{\citenamefont {Heusler}(1992)}]{Heusler:1992ss}%
  \BibitemOpen
  \bibfield  {author} {\bibinfo {author} {\bibfnamefont {M.}~\bibnamefont
  {Heusler}},\ }\href {\doibase 10.1063/1.529899} {\bibfield  {journal}
  {\bibinfo  {journal} {J. Math. Phys.}\ }\textbf {\bibinfo {volume} {33}},\
  \bibinfo {pages} {3497} (\bibinfo {year} {1992})}\BibitemShut {NoStop}%
\bibitem [{\citenamefont {Nunez}\ \emph {et~al.}(1996)\citenamefont {Nunez},
  \citenamefont {Quevedo},\ and\ \citenamefont {Sudarsky}}]{Nunez:1996xv}%
  \BibitemOpen
  \bibfield  {author} {\bibinfo {author} {\bibfnamefont {D.}~\bibnamefont
  {Nunez}}, \bibinfo {author} {\bibfnamefont {H.}~\bibnamefont {Quevedo}}, \
  and\ \bibinfo {author} {\bibfnamefont {D.}~\bibnamefont {Sudarsky}},\ }\href
  {\doibase 10.1103/PhysRevLett.76.571} {\bibfield  {journal} {\bibinfo
  {journal} {Phys. Rev. Lett.}\ }\textbf {\bibinfo {volume} {76}},\ \bibinfo
  {pages} {571} (\bibinfo {year} {1996})},\ \Eprint
  {http://arxiv.org/abs/gr-qc/9601020} {arXiv:gr-qc/9601020} \BibitemShut
  {NoStop}%
\bibitem [{\citenamefont {Banados}\ \emph {et~al.}(1992)\citenamefont
  {Banados}, \citenamefont {Teitelboim},\ and\ \citenamefont
  {Zanelli}}]{Banados:1992wn}%
  \BibitemOpen
  \bibfield  {author} {\bibinfo {author} {\bibfnamefont {M.}~\bibnamefont
  {Banados}}, \bibinfo {author} {\bibfnamefont {C.}~\bibnamefont {Teitelboim}},
  \ and\ \bibinfo {author} {\bibfnamefont {J.}~\bibnamefont {Zanelli}},\ }\href
  {\doibase 10.1103/PhysRevLett.69.1849} {\bibfield  {journal} {\bibinfo
  {journal} {Phys. Rev. Lett.}\ }\textbf {\bibinfo {volume} {69}},\ \bibinfo
  {pages} {1849} (\bibinfo {year} {1992})},\ \Eprint
  {http://arxiv.org/abs/hep-th/9204099} {arXiv:hep-th/9204099} \BibitemShut
  {NoStop}%
\bibitem [{\citenamefont {Banados}\ \emph {et~al.}(1993)\citenamefont
  {Banados}, \citenamefont {Henneaux}, \citenamefont {Teitelboim},\ and\
  \citenamefont {Zanelli}}]{Banados:1992gq}%
  \BibitemOpen
  \bibfield  {author} {\bibinfo {author} {\bibfnamefont {M.}~\bibnamefont
  {Banados}}, \bibinfo {author} {\bibfnamefont {M.}~\bibnamefont {Henneaux}},
  \bibinfo {author} {\bibfnamefont {C.}~\bibnamefont {Teitelboim}}, \ and\
  \bibinfo {author} {\bibfnamefont {J.}~\bibnamefont {Zanelli}},\ }\href
  {\doibase 10.1103/PhysRevD.48.1506} {\bibfield  {journal} {\bibinfo
  {journal} {Phys. Rev. D}\ }\textbf {\bibinfo {volume} {48}},\ \bibinfo
  {pages} {1506} (\bibinfo {year} {1993})},\ \bibinfo {note} {[Erratum:
  Phys.Rev.D 88, 069902 (2013)]},\ \Eprint {http://arxiv.org/abs/gr-qc/9302012}
  {arXiv:gr-qc/9302012} \BibitemShut {NoStop}%
\bibitem [{\citenamefont {Carlip}(1995)}]{Carlip:1995qv}%
  \BibitemOpen
  \bibfield  {author} {\bibinfo {author} {\bibfnamefont {S.}~\bibnamefont
  {Carlip}},\ }\href {\doibase 10.1088/0264-9381/12/12/005} {\bibfield
  {journal} {\bibinfo  {journal} {Class. Quant. Grav.}\ }\textbf {\bibinfo
  {volume} {12}},\ \bibinfo {pages} {2853} (\bibinfo {year} {1995})},\ \Eprint
  {http://arxiv.org/abs/gr-qc/9506079} {arXiv:gr-qc/9506079} \BibitemShut
  {NoStop}%
\bibitem [{\citenamefont {Strominger}(1998)}]{Strominger:1997eq}%
  \BibitemOpen
  \bibfield  {author} {\bibinfo {author} {\bibfnamefont {A.}~\bibnamefont
  {Strominger}},\ }\href {\doibase 10.1088/1126-6708/1998/02/009} {\bibfield
  {journal} {\bibinfo  {journal} {JHEP}\ }\textbf {\bibinfo {volume} {02}},\
  \bibinfo {pages} {009} (\bibinfo {year} {1998})},\ \Eprint
  {http://arxiv.org/abs/hep-th/9712251} {arXiv:hep-th/9712251} \BibitemShut
  {NoStop}%
\bibitem [{\citenamefont {Balthazar}\ \emph {et~al.}(2022)\citenamefont
  {Balthazar}, \citenamefont {Giveon}, \citenamefont {Kutasov},\ and\
  \citenamefont {Martinec}}]{Balthazar:2021xeh}%
  \BibitemOpen
  \bibfield  {author} {\bibinfo {author} {\bibfnamefont {B.}~\bibnamefont
  {Balthazar}}, \bibinfo {author} {\bibfnamefont {A.}~\bibnamefont {Giveon}},
  \bibinfo {author} {\bibfnamefont {D.}~\bibnamefont {Kutasov}}, \ and\
  \bibinfo {author} {\bibfnamefont {E.~J.}\ \bibnamefont {Martinec}},\ }\href
  {\doibase 10.1007/JHEP01(2022)008} {\bibfield  {journal} {\bibinfo  {journal}
  {JHEP}\ }\textbf {\bibinfo {volume} {01}},\ \bibinfo {pages} {008} (\bibinfo
  {year} {2022})},\ \Eprint {http://arxiv.org/abs/2109.00065} {arXiv:2109.00065
  [hep-th]} \BibitemShut {NoStop}%
\bibitem [{\citenamefont {Anninos}\ \emph {et~al.}(2009)\citenamefont
  {Anninos}, \citenamefont {Li}, \citenamefont {Padi}, \citenamefont {Song},\
  and\ \citenamefont {Strominger}}]{Anninos:2008fx}%
  \BibitemOpen
  \bibfield  {author} {\bibinfo {author} {\bibfnamefont {D.}~\bibnamefont
  {Anninos}}, \bibinfo {author} {\bibfnamefont {W.}~\bibnamefont {Li}},
  \bibinfo {author} {\bibfnamefont {M.}~\bibnamefont {Padi}}, \bibinfo {author}
  {\bibfnamefont {W.}~\bibnamefont {Song}}, \ and\ \bibinfo {author}
  {\bibfnamefont {A.}~\bibnamefont {Strominger}},\ }\href {\doibase
  10.1088/1126-6708/2009/03/130} {\bibfield  {journal} {\bibinfo  {journal}
  {JHEP}\ }\textbf {\bibinfo {volume} {03}},\ \bibinfo {pages} {130} (\bibinfo
  {year} {2009})},\ \Eprint {http://arxiv.org/abs/0807.3040} {arXiv:0807.3040
  [hep-th]} \BibitemShut {NoStop}%
\bibitem [{\citenamefont {Peet}(2000)}]{Peet:2000hn}%
  \BibitemOpen
  \bibfield  {author} {\bibinfo {author} {\bibfnamefont {A.~W.}\ \bibnamefont
  {Peet}}\ }(\bibinfo {year} {2000})\ pp.\ \bibinfo {pages} {353--433},\
  \Eprint {http://arxiv.org/abs/hep-th/0008241} {arXiv:hep-th/0008241}
  \BibitemShut {NoStop}%
\bibitem [{\citenamefont {Hartman}\ \emph {et~al.}(2014)\citenamefont
  {Hartman}, \citenamefont {Keller},\ and\ \citenamefont
  {Stoica}}]{Hartman:2014oaa}%
  \BibitemOpen
  \bibfield  {author} {\bibinfo {author} {\bibfnamefont {T.}~\bibnamefont
  {Hartman}}, \bibinfo {author} {\bibfnamefont {C.~A.}\ \bibnamefont {Keller}},
  \ and\ \bibinfo {author} {\bibfnamefont {B.}~\bibnamefont {Stoica}},\ }\href
  {\doibase 10.1007/JHEP09(2014)118} {\bibfield  {journal} {\bibinfo  {journal}
  {JHEP}\ }\textbf {\bibinfo {volume} {09}},\ \bibinfo {pages} {118} (\bibinfo
  {year} {2014})},\ \Eprint {http://arxiv.org/abs/1405.5137} {arXiv:1405.5137
  [hep-th]} \BibitemShut {NoStop}%
\bibitem [{\citenamefont {Correa}\ \emph {et~al.}(2012)\citenamefont {Correa},
  \citenamefont {Martinez},\ and\ \citenamefont {Troncoso}}]{Correa:2011dt}%
  \BibitemOpen
  \bibfield  {author} {\bibinfo {author} {\bibfnamefont {F.}~\bibnamefont
  {Correa}}, \bibinfo {author} {\bibfnamefont {C.}~\bibnamefont {Martinez}}, \
  and\ \bibinfo {author} {\bibfnamefont {R.}~\bibnamefont {Troncoso}},\ }\href
  {\doibase 10.1007/JHEP02(2012)136} {\bibfield  {journal} {\bibinfo  {journal}
  {JHEP}\ }\textbf {\bibinfo {volume} {02}},\ \bibinfo {pages} {136} (\bibinfo
  {year} {2012})},\ \Eprint {http://arxiv.org/abs/1112.6198} {arXiv:1112.6198
  [hep-th]} \BibitemShut {NoStop}%
\bibitem [{\citenamefont {Xu}\ and\ \citenamefont {Zou}(2017)}]{Xu:2014uka}%
  \BibitemOpen
  \bibfield  {author} {\bibinfo {author} {\bibfnamefont {W.}~\bibnamefont
  {Xu}}\ and\ \bibinfo {author} {\bibfnamefont {D.-C.}\ \bibnamefont {Zou}},\
  }\href {\doibase 10.1007/s10714-017-2237-4} {\bibfield  {journal} {\bibinfo
  {journal} {Gen. Rel. Grav.}\ }\textbf {\bibinfo {volume} {49}},\ \bibinfo
  {pages} {73} (\bibinfo {year} {2017})},\ \Eprint
  {http://arxiv.org/abs/1408.1998} {arXiv:1408.1998 [hep-th]} \BibitemShut
  {NoStop}%
\bibitem [{\citenamefont {Zhao}\ \emph {et~al.}(2014)\citenamefont {Zhao},
  \citenamefont {Xu},\ and\ \citenamefont {Zhu}}]{Zhao:2013isa}%
  \BibitemOpen
  \bibfield  {author} {\bibinfo {author} {\bibfnamefont {L.}~\bibnamefont
  {Zhao}}, \bibinfo {author} {\bibfnamefont {W.}~\bibnamefont {Xu}}, \ and\
  \bibinfo {author} {\bibfnamefont {B.}~\bibnamefont {Zhu}},\ }\href {\doibase
  10.1088/0253-6102/61/4/12} {\bibfield  {journal} {\bibinfo  {journal}
  {Commun. Theor. Phys.}\ }\textbf {\bibinfo {volume} {61}},\ \bibinfo {pages}
  {475} (\bibinfo {year} {2014})},\ \Eprint {http://arxiv.org/abs/1305.6001}
  {arXiv:1305.6001 [gr-qc]} \BibitemShut {NoStop}%
\bibitem [{\citenamefont {Perez}\ \emph {et~al.}(2013)\citenamefont {Perez},
  \citenamefont {Tempo},\ and\ \citenamefont {Troncoso}}]{Perez:2012cf}%
  \BibitemOpen
  \bibfield  {author} {\bibinfo {author} {\bibfnamefont {A.}~\bibnamefont
  {Perez}}, \bibinfo {author} {\bibfnamefont {D.}~\bibnamefont {Tempo}}, \ and\
  \bibinfo {author} {\bibfnamefont {R.}~\bibnamefont {Troncoso}},\ }\href
  {\doibase 10.1016/j.physletb.2013.08.038} {\bibfield  {journal} {\bibinfo
  {journal} {Phys. Lett. B}\ }\textbf {\bibinfo {volume} {726}},\ \bibinfo
  {pages} {444} (\bibinfo {year} {2013})},\ \Eprint
  {http://arxiv.org/abs/1207.2844} {arXiv:1207.2844 [hep-th]} \BibitemShut
  {NoStop}%
\bibitem [{\citenamefont {Bueno}\ \emph {et~al.}(2021)\citenamefont {Bueno},
  \citenamefont {Cano}, \citenamefont {Moreno},\ and\ \citenamefont {van~der
  Velde}}]{Bueno:2021krl}%
  \BibitemOpen
  \bibfield  {author} {\bibinfo {author} {\bibfnamefont {P.}~\bibnamefont
  {Bueno}}, \bibinfo {author} {\bibfnamefont {P.~A.}\ \bibnamefont {Cano}},
  \bibinfo {author} {\bibfnamefont {J.}~\bibnamefont {Moreno}}, \ and\ \bibinfo
  {author} {\bibfnamefont {G.}~\bibnamefont {van~der Velde}},\ }\href {\doibase
  10.1103/PhysRevD.104.L021501} {\bibfield  {journal} {\bibinfo  {journal}
  {Phys. Rev. D}\ }\textbf {\bibinfo {volume} {104}},\ \bibinfo {pages}
  {L021501} (\bibinfo {year} {2021})},\ \Eprint
  {http://arxiv.org/abs/2104.10172} {arXiv:2104.10172 [gr-qc]} \BibitemShut
  {NoStop}%
\bibitem [{\citenamefont {Greene}\ \emph {et~al.}(1993)\citenamefont {Greene},
  \citenamefont {Mathur},\ and\ \citenamefont {O'Neill}}]{Greene:1992fw}%
  \BibitemOpen
  \bibfield  {author} {\bibinfo {author} {\bibfnamefont {B.~R.}\ \bibnamefont
  {Greene}}, \bibinfo {author} {\bibfnamefont {S.~D.}\ \bibnamefont {Mathur}},
  \ and\ \bibinfo {author} {\bibfnamefont {C.~M.}\ \bibnamefont {O'Neill}},\
  }\href {\doibase 10.1103/PhysRevD.47.2242} {\bibfield  {journal} {\bibinfo
  {journal} {Phys. Rev. D}\ }\textbf {\bibinfo {volume} {47}},\ \bibinfo
  {pages} {2242} (\bibinfo {year} {1993})},\ \Eprint
  {http://arxiv.org/abs/hep-th/9211007} {arXiv:hep-th/9211007} \BibitemShut
  {NoStop}%
\bibitem [{\citenamefont {Bueno}\ and\ \citenamefont
  {Cano}(2017)}]{Bueno:2017sui}%
  \BibitemOpen
  \bibfield  {author} {\bibinfo {author} {\bibfnamefont {P.}~\bibnamefont
  {Bueno}}\ and\ \bibinfo {author} {\bibfnamefont {P.~A.}\ \bibnamefont
  {Cano}},\ }\href {\doibase 10.1088/1361-6382/aa8056} {\bibfield  {journal}
  {\bibinfo  {journal} {Class. Quant. Grav.}\ }\textbf {\bibinfo {volume}
  {34}},\ \bibinfo {pages} {175008} (\bibinfo {year} {2017})},\ \Eprint
  {http://arxiv.org/abs/1703.04625} {arXiv:1703.04625 [hep-th]} \BibitemShut
  {NoStop}%
\bibitem [{\citenamefont {Bueno}\ \emph {et~al.}(2022)\citenamefont {Bueno},
  \citenamefont {Cano}, \citenamefont {Llorens}, \citenamefont {Moreno},\ and\
  \citenamefont {van~der Velde}}]{Bueno:2022lhf}%
  \BibitemOpen
  \bibfield  {author} {\bibinfo {author} {\bibfnamefont {P.}~\bibnamefont
  {Bueno}}, \bibinfo {author} {\bibfnamefont {P.~A.}\ \bibnamefont {Cano}},
  \bibinfo {author} {\bibfnamefont {Q.}~\bibnamefont {Llorens}}, \bibinfo
  {author} {\bibfnamefont {J.}~\bibnamefont {Moreno}}, \ and\ \bibinfo {author}
  {\bibfnamefont {G.}~\bibnamefont {van~der Velde}},\ }\href {\doibase
  10.1088/1361-6382/ac6cbf} {\bibfield  {journal} {\bibinfo  {journal} {Class.
  Quant. Grav.}\ }\textbf {\bibinfo {volume} {39}},\ \bibinfo {pages} {125002}
  (\bibinfo {year} {2022})},\ \Eprint {http://arxiv.org/abs/2201.07266}
  {arXiv:2201.07266 [gr-qc]} \BibitemShut {NoStop}%
\bibitem [{\citenamefont {Cano}\ \emph {et~al.}(2022)\citenamefont {Cano},
  \citenamefont {Ganchev}, \citenamefont {Mayerson},\ and\ \citenamefont
  {Ruip\'erez}}]{Cano:2022wwo}%
  \BibitemOpen
  \bibfield  {author} {\bibinfo {author} {\bibfnamefont {P.~A.}\ \bibnamefont
  {Cano}}, \bibinfo {author} {\bibfnamefont {B.}~\bibnamefont {Ganchev}},
  \bibinfo {author} {\bibfnamefont {D.}~\bibnamefont {Mayerson}}, \ and\
  \bibinfo {author} {\bibfnamefont {A.}~\bibnamefont {Ruip\'erez}},\
  }\href@noop {} {\  (\bibinfo {year} {2022})},\ \Eprint
  {http://arxiv.org/abs/2208.01044} {arXiv:2208.01044 [gr-qc]} \BibitemShut
  {NoStop}%
\bibitem [{\citenamefont {Martinez}\ \emph {et~al.}(2006)\citenamefont
  {Martinez}, \citenamefont {Staforelli},\ and\ \citenamefont
  {Troncoso}}]{Martinez:2005di}%
  \BibitemOpen
  \bibfield  {author} {\bibinfo {author} {\bibfnamefont {C.}~\bibnamefont
  {Martinez}}, \bibinfo {author} {\bibfnamefont {J.~P.}\ \bibnamefont
  {Staforelli}}, \ and\ \bibinfo {author} {\bibfnamefont {R.}~\bibnamefont
  {Troncoso}},\ }\href {\doibase 10.1103/PhysRevD.74.044028} {\bibfield
  {journal} {\bibinfo  {journal} {Phys. Rev. D}\ }\textbf {\bibinfo {volume}
  {74}},\ \bibinfo {pages} {044028} (\bibinfo {year} {2006})},\ \Eprint
  {http://arxiv.org/abs/hep-th/0512022} {arXiv:hep-th/0512022} \BibitemShut
  {NoStop}%
\bibitem [{\citenamefont {Astefanesei}\ \emph
  {et~al.}(2020{\natexlab{a}})\citenamefont {Astefanesei}, \citenamefont
  {Bl\'azquez-Salcedo}, \citenamefont {Herdeiro}, \citenamefont {Radu},\ and\
  \citenamefont {Sanchis-Gual}}]{Astefanesei:2019qsg}%
  \BibitemOpen
  \bibfield  {author} {\bibinfo {author} {\bibfnamefont {D.}~\bibnamefont
  {Astefanesei}}, \bibinfo {author} {\bibfnamefont {J.~L.}\ \bibnamefont
  {Bl\'azquez-Salcedo}}, \bibinfo {author} {\bibfnamefont {C.}~\bibnamefont
  {Herdeiro}}, \bibinfo {author} {\bibfnamefont {E.}~\bibnamefont {Radu}}, \
  and\ \bibinfo {author} {\bibfnamefont {N.}~\bibnamefont {Sanchis-Gual}},\
  }\href {\doibase 10.1007/JHEP07(2020)063} {\bibfield  {journal} {\bibinfo
  {journal} {JHEP}\ }\textbf {\bibinfo {volume} {07}},\ \bibinfo {pages} {063}
  (\bibinfo {year} {2020}{\natexlab{a}})},\ \Eprint
  {http://arxiv.org/abs/1912.02192} {arXiv:1912.02192 [gr-qc]} \BibitemShut
  {NoStop}%
\bibitem [{\citenamefont {Anabal\'on}\ \emph {et~al.}(2022)\citenamefont
  {Anabal\'on}, \citenamefont {Concha}, \citenamefont {Oliva}, \citenamefont
  {Quijada},\ and\ \citenamefont {Rodr\'\i{}guez}}]{Anabalon:2022ksf}%
  \BibitemOpen
  \bibfield  {author} {\bibinfo {author} {\bibfnamefont {A.}~\bibnamefont
  {Anabal\'on}}, \bibinfo {author} {\bibfnamefont {P.}~\bibnamefont {Concha}},
  \bibinfo {author} {\bibfnamefont {J.}~\bibnamefont {Oliva}}, \bibinfo
  {author} {\bibfnamefont {C.}~\bibnamefont {Quijada}}, \ and\ \bibinfo
  {author} {\bibfnamefont {E.}~\bibnamefont {Rodr\'\i{}guez}},\ }\href@noop {}
  {\  (\bibinfo {year} {2022})},\ \Eprint {http://arxiv.org/abs/2205.01609}
  {arXiv:2205.01609 [hep-th]} \BibitemShut {NoStop}%
\bibitem [{\citenamefont {Anabalon}\ \emph {et~al.}(2021)\citenamefont
  {Anabalon}, \citenamefont {Astefanesei}, \citenamefont {Gallerati},\ and\
  \citenamefont {Trigiante}}]{Anabalon:2020pez}%
  \BibitemOpen
  \bibfield  {author} {\bibinfo {author} {\bibfnamefont {A.}~\bibnamefont
  {Anabalon}}, \bibinfo {author} {\bibfnamefont {D.}~\bibnamefont
  {Astefanesei}}, \bibinfo {author} {\bibfnamefont {A.}~\bibnamefont
  {Gallerati}}, \ and\ \bibinfo {author} {\bibfnamefont {M.}~\bibnamefont
  {Trigiante}},\ }\href {\doibase 10.1007/JHEP04(2021)047} {\bibfield
  {journal} {\bibinfo  {journal} {JHEP}\ }\textbf {\bibinfo {volume} {04}},\
  \bibinfo {pages} {047} (\bibinfo {year} {2021})},\ \Eprint
  {http://arxiv.org/abs/2012.09877} {arXiv:2012.09877 [hep-th]} \BibitemShut
  {NoStop}%
\bibitem [{\citenamefont {Gonz\'alez}\ \emph {et~al.}(2013)\citenamefont
  {Gonz\'alez}, \citenamefont {Papantonopoulos}, \citenamefont {Saavedra},\
  and\ \citenamefont {V\'asquez}}]{Gonzalez:2013aca}%
  \BibitemOpen
  \bibfield  {author} {\bibinfo {author} {\bibfnamefont {P.~A.}\ \bibnamefont
  {Gonz\'alez}}, \bibinfo {author} {\bibfnamefont {E.}~\bibnamefont
  {Papantonopoulos}}, \bibinfo {author} {\bibfnamefont {J.}~\bibnamefont
  {Saavedra}}, \ and\ \bibinfo {author} {\bibfnamefont {Y.}~\bibnamefont
  {V\'asquez}},\ }\href {\doibase 10.1007/JHEP12(2013)021} {\bibfield
  {journal} {\bibinfo  {journal} {JHEP}\ }\textbf {\bibinfo {volume} {12}},\
  \bibinfo {pages} {021} (\bibinfo {year} {2013})},\ \Eprint
  {http://arxiv.org/abs/1309.2161} {arXiv:1309.2161 [gr-qc]} \BibitemShut
  {NoStop}%
\bibitem [{\citenamefont {Martinez}\ \emph {et~al.}(2004)\citenamefont
  {Martinez}, \citenamefont {Troncoso},\ and\ \citenamefont
  {Zanelli}}]{Martinez:2004nb}%
  \BibitemOpen
  \bibfield  {author} {\bibinfo {author} {\bibfnamefont {C.}~\bibnamefont
  {Martinez}}, \bibinfo {author} {\bibfnamefont {R.}~\bibnamefont {Troncoso}},
  \ and\ \bibinfo {author} {\bibfnamefont {J.}~\bibnamefont {Zanelli}},\ }\href
  {\doibase 10.1103/PhysRevD.70.084035} {\bibfield  {journal} {\bibinfo
  {journal} {Phys. Rev. D}\ }\textbf {\bibinfo {volume} {70}},\ \bibinfo
  {pages} {084035} (\bibinfo {year} {2004})},\ \Eprint
  {http://arxiv.org/abs/hep-th/0406111} {arXiv:hep-th/0406111} \BibitemShut
  {NoStop}%
\bibitem [{\citenamefont {Martinez}\ \emph {et~al.}(2003)\citenamefont
  {Martinez}, \citenamefont {Troncoso},\ and\ \citenamefont
  {Zanelli}}]{Martinez:2002ru}%
  \BibitemOpen
  \bibfield  {author} {\bibinfo {author} {\bibfnamefont {C.}~\bibnamefont
  {Martinez}}, \bibinfo {author} {\bibfnamefont {R.}~\bibnamefont {Troncoso}},
  \ and\ \bibinfo {author} {\bibfnamefont {J.}~\bibnamefont {Zanelli}},\ }\href
  {\doibase 10.1103/PhysRevD.67.024008} {\bibfield  {journal} {\bibinfo
  {journal} {Phys. Rev. D}\ }\textbf {\bibinfo {volume} {67}},\ \bibinfo
  {pages} {024008} (\bibinfo {year} {2003})},\ \Eprint
  {http://arxiv.org/abs/hep-th/0205319} {arXiv:hep-th/0205319} \BibitemShut
  {NoStop}%
\bibitem [{\citenamefont {Anabal\'on}\ and\ \citenamefont
  {Astefanesei}(2013)}]{Anabalon:2013sra}%
  \BibitemOpen
  \bibfield  {author} {\bibinfo {author} {\bibfnamefont {A.}~\bibnamefont
  {Anabal\'on}}\ and\ \bibinfo {author} {\bibfnamefont {D.}~\bibnamefont
  {Astefanesei}},\ }\href {\doibase 10.1016/j.physletb.2013.11.013} {\bibfield
  {journal} {\bibinfo  {journal} {Phys. Lett. B}\ }\textbf {\bibinfo {volume}
  {727}},\ \bibinfo {pages} {568} (\bibinfo {year} {2013})},\ \Eprint
  {http://arxiv.org/abs/1309.5863} {arXiv:1309.5863 [hep-th]} \BibitemShut
  {NoStop}%
\bibitem [{\citenamefont {Anabalon}\ \emph {et~al.}(2013)\citenamefont
  {Anabalon}, \citenamefont {Astefanesei},\ and\ \citenamefont
  {Mann}}]{Anabalon:2013qua}%
  \BibitemOpen
  \bibfield  {author} {\bibinfo {author} {\bibfnamefont {A.}~\bibnamefont
  {Anabalon}}, \bibinfo {author} {\bibfnamefont {D.}~\bibnamefont
  {Astefanesei}}, \ and\ \bibinfo {author} {\bibfnamefont {R.}~\bibnamefont
  {Mann}},\ }\href {\doibase 10.1007/JHEP10(2013)184} {\bibfield  {journal}
  {\bibinfo  {journal} {JHEP}\ }\textbf {\bibinfo {volume} {10}},\ \bibinfo
  {pages} {184} (\bibinfo {year} {2013})},\ \Eprint
  {http://arxiv.org/abs/1308.1693} {arXiv:1308.1693 [hep-th]} \BibitemShut
  {NoStop}%
\bibitem [{\citenamefont {Acena}\ \emph {et~al.}(2013)\citenamefont {Acena},
  \citenamefont {Anabalon},\ and\ \citenamefont {Astefanesei}}]{Acena:2012mr}%
  \BibitemOpen
  \bibfield  {author} {\bibinfo {author} {\bibfnamefont {A.}~\bibnamefont
  {Acena}}, \bibinfo {author} {\bibfnamefont {A.}~\bibnamefont {Anabalon}}, \
  and\ \bibinfo {author} {\bibfnamefont {D.}~\bibnamefont {Astefanesei}},\
  }\href {\doibase 10.1103/PhysRevD.87.124033} {\bibfield  {journal} {\bibinfo
  {journal} {Phys. Rev. D}\ }\textbf {\bibinfo {volume} {87}},\ \bibinfo
  {pages} {124033} (\bibinfo {year} {2013})},\ \Eprint
  {http://arxiv.org/abs/1211.6126} {arXiv:1211.6126 [hep-th]} \BibitemShut
  {NoStop}%
\bibitem [{\citenamefont {Ace\~na}\ \emph {et~al.}(2014)\citenamefont
  {Ace\~na}, \citenamefont {Anabal\'on}, \citenamefont {Astefanesei},\ and\
  \citenamefont {Mann}}]{Acena:2013jya}%
  \BibitemOpen
  \bibfield  {author} {\bibinfo {author} {\bibfnamefont {A.}~\bibnamefont
  {Ace\~na}}, \bibinfo {author} {\bibfnamefont {A.}~\bibnamefont {Anabal\'on}},
  \bibinfo {author} {\bibfnamefont {D.}~\bibnamefont {Astefanesei}}, \ and\
  \bibinfo {author} {\bibfnamefont {R.}~\bibnamefont {Mann}},\ }\href {\doibase
  10.1007/JHEP01(2014)153} {\bibfield  {journal} {\bibinfo  {journal} {JHEP}\
  }\textbf {\bibinfo {volume} {01}},\ \bibinfo {pages} {153} (\bibinfo {year}
  {2014})},\ \Eprint {http://arxiv.org/abs/1311.6065} {arXiv:1311.6065
  [hep-th]} \BibitemShut {NoStop}%
\bibitem [{\citenamefont {Hawking}\ and\ \citenamefont
  {Page}(1983)}]{Hawking:1982dh}%
  \BibitemOpen
  \bibfield  {author} {\bibinfo {author} {\bibfnamefont {S.~W.}\ \bibnamefont
  {Hawking}}\ and\ \bibinfo {author} {\bibfnamefont {D.~N.}\ \bibnamefont
  {Page}},\ }\href {\doibase 10.1007/BF01208266} {\bibfield  {journal}
  {\bibinfo  {journal} {Commun. Math. Phys.}\ }\textbf {\bibinfo {volume}
  {87}},\ \bibinfo {pages} {577} (\bibinfo {year} {1983})}\BibitemShut
  {NoStop}%
\bibitem [{\citenamefont {Witten}(1998{\natexlab{a}})}]{Witten:1998zw}%
  \BibitemOpen
  \bibfield  {author} {\bibinfo {author} {\bibfnamefont {E.}~\bibnamefont
  {Witten}},\ }\href {\doibase 10.4310/ATMP.1998.v2.n3.a3} {\bibfield
  {journal} {\bibinfo  {journal} {Adv. Theor. Math. Phys.}\ }\textbf {\bibinfo
  {volume} {2}},\ \bibinfo {pages} {505} (\bibinfo {year}
  {1998}{\natexlab{a}})},\ \Eprint {http://arxiv.org/abs/hep-th/9803131}
  {arXiv:hep-th/9803131} \BibitemShut {NoStop}%
\bibitem [{\citenamefont {Aharony}\ \emph {et~al.}(2000)\citenamefont
  {Aharony}, \citenamefont {Gubser}, \citenamefont {Maldacena}, \citenamefont
  {Ooguri},\ and\ \citenamefont {Oz}}]{Aharony:1999ti}%
  \BibitemOpen
  \bibfield  {author} {\bibinfo {author} {\bibfnamefont {O.}~\bibnamefont
  {Aharony}}, \bibinfo {author} {\bibfnamefont {S.~S.}\ \bibnamefont {Gubser}},
  \bibinfo {author} {\bibfnamefont {J.~M.}\ \bibnamefont {Maldacena}}, \bibinfo
  {author} {\bibfnamefont {H.}~\bibnamefont {Ooguri}}, \ and\ \bibinfo {author}
  {\bibfnamefont {Y.}~\bibnamefont {Oz}},\ }\href {\doibase
  10.1016/S0370-1573(99)00083-6} {\bibfield  {journal} {\bibinfo  {journal}
  {Phys. Rept.}\ }\textbf {\bibinfo {volume} {323}},\ \bibinfo {pages} {183}
  (\bibinfo {year} {2000})},\ \Eprint {http://arxiv.org/abs/hep-th/9905111}
  {arXiv:hep-th/9905111} \BibitemShut {NoStop}%
\bibitem [{\citenamefont {Gubser}\ \emph {et~al.}(1998)\citenamefont {Gubser},
  \citenamefont {Klebanov},\ and\ \citenamefont {Polyakov}}]{Gubser:1998bc}%
  \BibitemOpen
  \bibfield  {author} {\bibinfo {author} {\bibfnamefont {S.~S.}\ \bibnamefont
  {Gubser}}, \bibinfo {author} {\bibfnamefont {I.~R.}\ \bibnamefont
  {Klebanov}}, \ and\ \bibinfo {author} {\bibfnamefont {A.~M.}\ \bibnamefont
  {Polyakov}},\ }\href {\doibase 10.1016/S0370-2693(98)00377-3} {\bibfield
  {journal} {\bibinfo  {journal} {Phys. Lett. B}\ }\textbf {\bibinfo {volume}
  {428}},\ \bibinfo {pages} {105} (\bibinfo {year} {1998})},\ \Eprint
  {http://arxiv.org/abs/hep-th/9802109} {arXiv:hep-th/9802109} \BibitemShut
  {NoStop}%
\bibitem [{\citenamefont {Witten}(1998{\natexlab{b}})}]{Witten:1998qj}%
  \BibitemOpen
  \bibfield  {author} {\bibinfo {author} {\bibfnamefont {E.}~\bibnamefont
  {Witten}},\ }\href {\doibase 10.4310/ATMP.1998.v2.n2.a2} {\bibfield
  {journal} {\bibinfo  {journal} {Adv. Theor. Math. Phys.}\ }\textbf {\bibinfo
  {volume} {2}},\ \bibinfo {pages} {253} (\bibinfo {year}
  {1998}{\natexlab{b}})},\ \Eprint {http://arxiv.org/abs/hep-th/9802150}
  {arXiv:hep-th/9802150} \BibitemShut {NoStop}%
\bibitem [{\citenamefont {Anabalon}\ \emph {et~al.}(2020)\citenamefont
  {Anabalon}, \citenamefont {Astefanesei}, \citenamefont {Choque},\ and\
  \citenamefont {Edelstein}}]{Anabalon:2019tcy}%
  \BibitemOpen
  \bibfield  {author} {\bibinfo {author} {\bibfnamefont {A.}~\bibnamefont
  {Anabalon}}, \bibinfo {author} {\bibfnamefont {D.}~\bibnamefont
  {Astefanesei}}, \bibinfo {author} {\bibfnamefont {D.}~\bibnamefont {Choque}},
  \ and\ \bibinfo {author} {\bibfnamefont {J.~D.}\ \bibnamefont {Edelstein}},\
  }\href {\doibase 10.1007/JHEP07(2020)129} {\bibfield  {journal} {\bibinfo
  {journal} {JHEP}\ }\textbf {\bibinfo {volume} {07}},\ \bibinfo {pages} {129}
  (\bibinfo {year} {2020})},\ \Eprint {http://arxiv.org/abs/1912.03318}
  {arXiv:1912.03318 [hep-th]} \BibitemShut {NoStop}%
\bibitem [{\citenamefont {Anabalon}\ \emph {et~al.}(2015)\citenamefont
  {Anabalon}, \citenamefont {Astefanesei},\ and\ \citenamefont
  {Choque}}]{Anabalon:2015ija}%
  \BibitemOpen
  \bibfield  {author} {\bibinfo {author} {\bibfnamefont {A.}~\bibnamefont
  {Anabalon}}, \bibinfo {author} {\bibfnamefont {D.}~\bibnamefont
  {Astefanesei}}, \ and\ \bibinfo {author} {\bibfnamefont {D.}~\bibnamefont
  {Choque}},\ }\href {\doibase 10.1016/j.physletb.2015.02.024} {\bibfield
  {journal} {\bibinfo  {journal} {Phys. Lett. B}\ }\textbf {\bibinfo {volume}
  {743}},\ \bibinfo {pages} {154} (\bibinfo {year} {2015})},\ \Eprint
  {http://arxiv.org/abs/1501.04252} {arXiv:1501.04252 [hep-th]} \BibitemShut
  {NoStop}%
\bibitem [{\citenamefont {Chamblin}\ \emph {et~al.}(1999)\citenamefont
  {Chamblin}, \citenamefont {Emparan}, \citenamefont {Johnson},\ and\
  \citenamefont {Myers}}]{Chamblin:1999hg}%
  \BibitemOpen
  \bibfield  {author} {\bibinfo {author} {\bibfnamefont {A.}~\bibnamefont
  {Chamblin}}, \bibinfo {author} {\bibfnamefont {R.}~\bibnamefont {Emparan}},
  \bibinfo {author} {\bibfnamefont {C.~V.}\ \bibnamefont {Johnson}}, \ and\
  \bibinfo {author} {\bibfnamefont {R.~C.}\ \bibnamefont {Myers}},\ }\href
  {\doibase 10.1103/PhysRevD.60.104026} {\bibfield  {journal} {\bibinfo
  {journal} {Phys. Rev. D}\ }\textbf {\bibinfo {volume} {60}},\ \bibinfo
  {pages} {104026} (\bibinfo {year} {1999})},\ \Eprint
  {http://arxiv.org/abs/hep-th/9904197} {arXiv:hep-th/9904197} \BibitemShut
  {NoStop}%
\bibitem [{\citenamefont {L\"u}\ \emph {et~al.}(2013)\citenamefont {L\"u},
  \citenamefont {Pang},\ and\ \citenamefont {Pope}}]{Lu:2013ura}%
  \BibitemOpen
  \bibfield  {author} {\bibinfo {author} {\bibfnamefont {H.}~\bibnamefont
  {L\"u}}, \bibinfo {author} {\bibfnamefont {Y.}~\bibnamefont {Pang}}, \ and\
  \bibinfo {author} {\bibfnamefont {C.~N.}\ \bibnamefont {Pope}},\ }\href
  {\doibase 10.1007/JHEP11(2013)033} {\bibfield  {journal} {\bibinfo  {journal}
  {JHEP}\ }\textbf {\bibinfo {volume} {11}},\ \bibinfo {pages} {033} (\bibinfo
  {year} {2013})},\ \Eprint {http://arxiv.org/abs/1307.6243} {arXiv:1307.6243
  [hep-th]} \BibitemShut {NoStop}%
\bibitem [{\citenamefont {Giribet}\ \emph {et~al.}(2015)\citenamefont
  {Giribet}, \citenamefont {Goya},\ and\ \citenamefont
  {Oliva}}]{Giribet:2014fla}%
  \BibitemOpen
  \bibfield  {author} {\bibinfo {author} {\bibfnamefont {G.}~\bibnamefont
  {Giribet}}, \bibinfo {author} {\bibfnamefont {A.}~\bibnamefont {Goya}}, \
  and\ \bibinfo {author} {\bibfnamefont {J.}~\bibnamefont {Oliva}},\ }\href
  {\doibase 10.1103/PhysRevD.91.045031} {\bibfield  {journal} {\bibinfo
  {journal} {Phys. Rev. D}\ }\textbf {\bibinfo {volume} {91}},\ \bibinfo
  {pages} {045031} (\bibinfo {year} {2015})},\ \Eprint
  {http://arxiv.org/abs/1501.00184} {arXiv:1501.00184 [hep-th]} \BibitemShut
  {NoStop}%
\bibitem [{\citenamefont {Astefanesei}\ \emph
  {et~al.}(2020{\natexlab{b}})\citenamefont {Astefanesei}, \citenamefont
  {Bl\'azquez-Salcedo}, \citenamefont {G\'omez},\ and\ \citenamefont
  {Rojas}}]{Astefanesei:2020xvn}%
  \BibitemOpen
  \bibfield  {author} {\bibinfo {author} {\bibfnamefont {D.}~\bibnamefont
  {Astefanesei}}, \bibinfo {author} {\bibfnamefont {J.~L.}\ \bibnamefont
  {Bl\'azquez-Salcedo}}, \bibinfo {author} {\bibfnamefont {F.}~\bibnamefont
  {G\'omez}}, \ and\ \bibinfo {author} {\bibfnamefont {R.}~\bibnamefont
  {Rojas}},\ }\href@noop {} {\  (\bibinfo {year} {2020}{\natexlab{b}})},\
  \Eprint {http://arxiv.org/abs/2009.01854} {arXiv:2009.01854 [hep-th]}
  \BibitemShut {NoStop}%
\bibitem [{\citenamefont {Astefanesei}\ \emph
  {et~al.}(2019{\natexlab{a}})\citenamefont {Astefanesei}, \citenamefont
  {Choque}, \citenamefont {G\'omez},\ and\ \citenamefont
  {Rojas}}]{Astefanesei:2019mds}%
  \BibitemOpen
  \bibfield  {author} {\bibinfo {author} {\bibfnamefont {D.}~\bibnamefont
  {Astefanesei}}, \bibinfo {author} {\bibfnamefont {D.}~\bibnamefont {Choque}},
  \bibinfo {author} {\bibfnamefont {F.}~\bibnamefont {G\'omez}}, \ and\
  \bibinfo {author} {\bibfnamefont {R.}~\bibnamefont {Rojas}},\ }\href
  {\doibase 10.1007/JHEP03(2019)205} {\bibfield  {journal} {\bibinfo  {journal}
  {JHEP}\ }\textbf {\bibinfo {volume} {03}},\ \bibinfo {pages} {205} (\bibinfo
  {year} {2019}{\natexlab{a}})},\ \Eprint {http://arxiv.org/abs/1901.01269}
  {arXiv:1901.01269 [hep-th]} \BibitemShut {NoStop}%
\bibitem [{\citenamefont {Kubiznak}\ and\ \citenamefont
  {Mann}(2012)}]{Kubiznak:2012wp}%
  \BibitemOpen
  \bibfield  {author} {\bibinfo {author} {\bibfnamefont {D.}~\bibnamefont
  {Kubiznak}}\ and\ \bibinfo {author} {\bibfnamefont {R.~B.}\ \bibnamefont
  {Mann}},\ }\href {\doibase 10.1007/JHEP07(2012)033} {\bibfield  {journal}
  {\bibinfo  {journal} {JHEP}\ }\textbf {\bibinfo {volume} {07}},\ \bibinfo
  {pages} {033} (\bibinfo {year} {2012})},\ \Eprint
  {http://arxiv.org/abs/1205.0559} {arXiv:1205.0559 [hep-th]} \BibitemShut
  {NoStop}%
\bibitem [{\citenamefont {Gunasekaran}\ \emph {et~al.}(2012)\citenamefont
  {Gunasekaran}, \citenamefont {Mann},\ and\ \citenamefont
  {Kubiznak}}]{Gunasekaran:2012dq}%
  \BibitemOpen
  \bibfield  {author} {\bibinfo {author} {\bibfnamefont {S.}~\bibnamefont
  {Gunasekaran}}, \bibinfo {author} {\bibfnamefont {R.~B.}\ \bibnamefont
  {Mann}}, \ and\ \bibinfo {author} {\bibfnamefont {D.}~\bibnamefont
  {Kubiznak}},\ }\href {\doibase 10.1007/JHEP11(2012)110} {\bibfield  {journal}
  {\bibinfo  {journal} {JHEP}\ }\textbf {\bibinfo {volume} {11}},\ \bibinfo
  {pages} {110} (\bibinfo {year} {2012})},\ \Eprint
  {http://arxiv.org/abs/1208.6251} {arXiv:1208.6251 [hep-th]} \BibitemShut
  {NoStop}%
\bibitem [{\citenamefont {Astefanesei}\ \emph
  {et~al.}(2019{\natexlab{b}})\citenamefont {Astefanesei}, \citenamefont
  {Mann},\ and\ \citenamefont {Rojas}}]{Astefanesei:2019ehu}%
  \BibitemOpen
  \bibfield  {author} {\bibinfo {author} {\bibfnamefont {D.}~\bibnamefont
  {Astefanesei}}, \bibinfo {author} {\bibfnamefont {R.~B.}\ \bibnamefont
  {Mann}}, \ and\ \bibinfo {author} {\bibfnamefont {R.}~\bibnamefont {Rojas}},\
  }\href {\doibase 10.1007/JHEP11(2019)043} {\bibfield  {journal} {\bibinfo
  {journal} {JHEP}\ }\textbf {\bibinfo {volume} {11}},\ \bibinfo {pages} {043}
  (\bibinfo {year} {2019}{\natexlab{b}})},\ \Eprint
  {http://arxiv.org/abs/1907.08636} {arXiv:1907.08636 [hep-th]} \BibitemShut
  {NoStop}%
\bibitem [{\citenamefont {Astefanesei}\ \emph {et~al.}(2018)\citenamefont
  {Astefanesei}, \citenamefont {Ballesteros}, \citenamefont {Choque},\ and\
  \citenamefont {Rojas}}]{Astefanesei:2018vga}%
  \BibitemOpen
  \bibfield  {author} {\bibinfo {author} {\bibfnamefont {D.}~\bibnamefont
  {Astefanesei}}, \bibinfo {author} {\bibfnamefont {R.}~\bibnamefont
  {Ballesteros}}, \bibinfo {author} {\bibfnamefont {D.}~\bibnamefont {Choque}},
  \ and\ \bibinfo {author} {\bibfnamefont {R.}~\bibnamefont {Rojas}},\ }\href
  {\doibase 10.1016/j.physletb.2018.05.005} {\bibfield  {journal} {\bibinfo
  {journal} {Phys. Lett. B}\ }\textbf {\bibinfo {volume} {782}},\ \bibinfo
  {pages} {47} (\bibinfo {year} {2018})},\ \Eprint
  {http://arxiv.org/abs/1803.11317} {arXiv:1803.11317 [hep-th]} \BibitemShut
  {NoStop}%
\bibitem [{\citenamefont {Anabalon}\ and\ \citenamefont
  {Astefanesei}(2014)}]{Anabalon:2013eaa}%
  \BibitemOpen
  \bibfield  {author} {\bibinfo {author} {\bibfnamefont {A.}~\bibnamefont
  {Anabalon}}\ and\ \bibinfo {author} {\bibfnamefont {D.}~\bibnamefont
  {Astefanesei}},\ }\href {\doibase 10.1016/j.physletb.2014.03.035} {\bibfield
  {journal} {\bibinfo  {journal} {Phys. Lett. B}\ }\textbf {\bibinfo {volume}
  {732}},\ \bibinfo {pages} {137} (\bibinfo {year} {2014})},\ \Eprint
  {http://arxiv.org/abs/1311.7459} {arXiv:1311.7459 [hep-th]} \BibitemShut
  {NoStop}%
\bibitem [{\citenamefont {Anabal\'on}\ \emph {et~al.}(2018)\citenamefont
  {Anabal\'on}, \citenamefont {Astefanesei}, \citenamefont {Gallerati},\ and\
  \citenamefont {Trigiante}}]{Anabalon:2017yhv}%
  \BibitemOpen
  \bibfield  {author} {\bibinfo {author} {\bibfnamefont {A.}~\bibnamefont
  {Anabal\'on}}, \bibinfo {author} {\bibfnamefont {D.}~\bibnamefont
  {Astefanesei}}, \bibinfo {author} {\bibfnamefont {A.}~\bibnamefont
  {Gallerati}}, \ and\ \bibinfo {author} {\bibfnamefont {M.}~\bibnamefont
  {Trigiante}},\ }\href {\doibase 10.1007/JHEP04(2018)058} {\bibfield
  {journal} {\bibinfo  {journal} {JHEP}\ }\textbf {\bibinfo {volume} {04}},\
  \bibinfo {pages} {058} (\bibinfo {year} {2018})},\ \Eprint
  {http://arxiv.org/abs/1712.06971} {arXiv:1712.06971 [hep-th]} \BibitemShut
  {NoStop}%
\bibitem [{\citenamefont {Anabal\'on}\ \emph {et~al.}(2021)\citenamefont
  {Anabal\'on}, \citenamefont {Astefanesei}, \citenamefont {Choque},
  \citenamefont {Gallerati},\ and\ \citenamefont
  {Trigiante}}]{Anabalon:2020qux}%
  \BibitemOpen
  \bibfield  {author} {\bibinfo {author} {\bibfnamefont {A.}~\bibnamefont
  {Anabal\'on}}, \bibinfo {author} {\bibfnamefont {D.}~\bibnamefont
  {Astefanesei}}, \bibinfo {author} {\bibfnamefont {D.}~\bibnamefont {Choque}},
  \bibinfo {author} {\bibfnamefont {A.}~\bibnamefont {Gallerati}}, \ and\
  \bibinfo {author} {\bibfnamefont {M.}~\bibnamefont {Trigiante}},\ }\href
  {\doibase 10.1007/JHEP04(2021)053} {\bibfield  {journal} {\bibinfo  {journal}
  {JHEP}\ }\textbf {\bibinfo {volume} {04}},\ \bibinfo {pages} {053} (\bibinfo
  {year} {2021})},\ \Eprint {http://arxiv.org/abs/2012.01289} {arXiv:2012.01289
  [hep-th]} \BibitemShut {NoStop}%
\bibitem [{\citenamefont {Henneaux}\ \emph {et~al.}(2002)\citenamefont
  {Henneaux}, \citenamefont {Martinez}, \citenamefont {Troncoso},\ and\
  \citenamefont {Zanelli}}]{Henneaux:2002wm}%
  \BibitemOpen
  \bibfield  {author} {\bibinfo {author} {\bibfnamefont {M.}~\bibnamefont
  {Henneaux}}, \bibinfo {author} {\bibfnamefont {C.}~\bibnamefont {Martinez}},
  \bibinfo {author} {\bibfnamefont {R.}~\bibnamefont {Troncoso}}, \ and\
  \bibinfo {author} {\bibfnamefont {J.}~\bibnamefont {Zanelli}},\ }\href
  {\doibase 10.1103/PhysRevD.65.104007} {\bibfield  {journal} {\bibinfo
  {journal} {Phys. Rev. D}\ }\textbf {\bibinfo {volume} {65}},\ \bibinfo
  {pages} {104007} (\bibinfo {year} {2002})},\ \Eprint
  {http://arxiv.org/abs/hep-th/0201170} {arXiv:hep-th/0201170} \BibitemShut
  {NoStop}%
\bibitem [{\citenamefont {Tang}\ \emph {et~al.}(2019)\citenamefont {Tang},
  \citenamefont {Ong}, \citenamefont {Wang},\ and\ \citenamefont
  {Papantonopoulos}}]{Tang:2019jkn}%
  \BibitemOpen
  \bibfield  {author} {\bibinfo {author} {\bibfnamefont {Z.-Y.}\ \bibnamefont
  {Tang}}, \bibinfo {author} {\bibfnamefont {Y.~C.}\ \bibnamefont {Ong}},
  \bibinfo {author} {\bibfnamefont {B.}~\bibnamefont {Wang}}, \ and\ \bibinfo
  {author} {\bibfnamefont {E.}~\bibnamefont {Papantonopoulos}},\ }\href
  {\doibase 10.1103/PhysRevD.100.024003} {\bibfield  {journal} {\bibinfo
  {journal} {Phys. Rev. D}\ }\textbf {\bibinfo {volume} {100}},\ \bibinfo
  {pages} {024003} (\bibinfo {year} {2019})},\ \Eprint
  {http://arxiv.org/abs/1901.07310} {arXiv:1901.07310 [gr-qc]} \BibitemShut
  {NoStop}%
\bibitem [{\citenamefont {Correa}\ \emph {et~al.}(2011)\citenamefont {Correa},
  \citenamefont {Martinez},\ and\ \citenamefont {Troncoso}}]{Correa:2010hf}%
  \BibitemOpen
  \bibfield  {author} {\bibinfo {author} {\bibfnamefont {F.}~\bibnamefont
  {Correa}}, \bibinfo {author} {\bibfnamefont {C.}~\bibnamefont {Martinez}}, \
  and\ \bibinfo {author} {\bibfnamefont {R.}~\bibnamefont {Troncoso}},\ }\href
  {\doibase 10.1007/JHEP01(2011)034} {\bibfield  {journal} {\bibinfo  {journal}
  {JHEP}\ }\textbf {\bibinfo {volume} {01}},\ \bibinfo {pages} {034} (\bibinfo
  {year} {2011})},\ \Eprint {http://arxiv.org/abs/1010.1259} {arXiv:1010.1259
  [hep-th]} \BibitemShut {NoStop}%
\bibitem [{\citenamefont {Anabalon}\ \emph
  {et~al.}(2016{\natexlab{a}})\citenamefont {Anabalon}, \citenamefont
  {Astefanesei}, \citenamefont {Choque},\ and\ \citenamefont
  {Martinez}}]{Anabalon:2015xvl}%
  \BibitemOpen
  \bibfield  {author} {\bibinfo {author} {\bibfnamefont {A.}~\bibnamefont
  {Anabalon}}, \bibinfo {author} {\bibfnamefont {D.}~\bibnamefont
  {Astefanesei}}, \bibinfo {author} {\bibfnamefont {D.}~\bibnamefont {Choque}},
  \ and\ \bibinfo {author} {\bibfnamefont {C.}~\bibnamefont {Martinez}},\
  }\href {\doibase 10.1007/JHEP03(2016)117} {\bibfield  {journal} {\bibinfo
  {journal} {JHEP}\ }\textbf {\bibinfo {volume} {03}},\ \bibinfo {pages} {117}
  (\bibinfo {year} {2016}{\natexlab{a}})},\ \Eprint
  {http://arxiv.org/abs/1511.08759} {arXiv:1511.08759 [hep-th]} \BibitemShut
  {NoStop}%
\bibitem [{\citenamefont {Hertog}\ and\ \citenamefont
  {Maeda}(2004)}]{Hertog:2004dr}%
  \BibitemOpen
  \bibfield  {author} {\bibinfo {author} {\bibfnamefont {T.}~\bibnamefont
  {Hertog}}\ and\ \bibinfo {author} {\bibfnamefont {K.}~\bibnamefont {Maeda}},\
  }\href {\doibase 10.1088/1126-6708/2004/07/051} {\bibfield  {journal}
  {\bibinfo  {journal} {JHEP}\ }\textbf {\bibinfo {volume} {07}},\ \bibinfo
  {pages} {051} (\bibinfo {year} {2004})},\ \Eprint
  {http://arxiv.org/abs/hep-th/0404261} {arXiv:hep-th/0404261} \BibitemShut
  {NoStop}%
\bibitem [{\citenamefont {Gallerati}(2021)}]{Gallerati:2021cty}%
  \BibitemOpen
  \bibfield  {author} {\bibinfo {author} {\bibfnamefont {A.}~\bibnamefont
  {Gallerati}},\ }\href {\doibase 10.3390/universe7060187} {\bibfield
  {journal} {\bibinfo  {journal} {Universe}\ }\textbf {\bibinfo {volume} {7}},\
  \bibinfo {pages} {187} (\bibinfo {year} {2021})}\BibitemShut {NoStop}%
\bibitem [{\citenamefont {Papadimitriou}(2007)}]{Papadimitriou:2007sj}%
  \BibitemOpen
  \bibfield  {author} {\bibinfo {author} {\bibfnamefont {I.}~\bibnamefont
  {Papadimitriou}},\ }\href {\doibase 10.1088/1126-6708/2007/05/075} {\bibfield
   {journal} {\bibinfo  {journal} {JHEP}\ }\textbf {\bibinfo {volume} {05}},\
  \bibinfo {pages} {075} (\bibinfo {year} {2007})},\ \Eprint
  {http://arxiv.org/abs/hep-th/0703152} {arXiv:hep-th/0703152} \BibitemShut
  {NoStop}%
\bibitem [{\citenamefont {Batrachenko}\ \emph {et~al.}(2005)\citenamefont
  {Batrachenko}, \citenamefont {Liu}, \citenamefont {McNees}, \citenamefont
  {Sabra},\ and\ \citenamefont {Wen}}]{Batrachenko:2004fd}%
  \BibitemOpen
  \bibfield  {author} {\bibinfo {author} {\bibfnamefont {A.}~\bibnamefont
  {Batrachenko}}, \bibinfo {author} {\bibfnamefont {J.~T.}\ \bibnamefont
  {Liu}}, \bibinfo {author} {\bibfnamefont {R.}~\bibnamefont {McNees}},
  \bibinfo {author} {\bibfnamefont {W.~A.}\ \bibnamefont {Sabra}}, \ and\
  \bibinfo {author} {\bibfnamefont {W.~Y.}\ \bibnamefont {Wen}},\ }\href
  {\doibase 10.1088/1126-6708/2005/05/034} {\bibfield  {journal} {\bibinfo
  {journal} {JHEP}\ }\textbf {\bibinfo {volume} {05}},\ \bibinfo {pages} {034}
  (\bibinfo {year} {2005})},\ \Eprint {http://arxiv.org/abs/hep-th/0408205}
  {arXiv:hep-th/0408205} \BibitemShut {NoStop}%
\bibitem [{\citenamefont {Gursoy}\ \emph {et~al.}(2009)\citenamefont {Gursoy},
  \citenamefont {Kiritsis}, \citenamefont {Mazzanti},\ and\ \citenamefont
  {Nitti}}]{Gursoy:2008za}%
  \BibitemOpen
  \bibfield  {author} {\bibinfo {author} {\bibfnamefont {U.}~\bibnamefont
  {Gursoy}}, \bibinfo {author} {\bibfnamefont {E.}~\bibnamefont {Kiritsis}},
  \bibinfo {author} {\bibfnamefont {L.}~\bibnamefont {Mazzanti}}, \ and\
  \bibinfo {author} {\bibfnamefont {F.}~\bibnamefont {Nitti}},\ }\href
  {\doibase 10.1088/1126-6708/2009/05/033} {\bibfield  {journal} {\bibinfo
  {journal} {JHEP}\ }\textbf {\bibinfo {volume} {05}},\ \bibinfo {pages} {033}
  (\bibinfo {year} {2009})},\ \Eprint {http://arxiv.org/abs/0812.0792}
  {arXiv:0812.0792 [hep-th]} \BibitemShut {NoStop}%
\bibitem [{\citenamefont {Astefanesei}\ \emph {et~al.}(2022)\citenamefont
  {Astefanesei}, \citenamefont {Choque}, \citenamefont {Maggiolo},\ and\
  \citenamefont {Rojas}}]{Astefanesei:2021ryn}%
  \BibitemOpen
  \bibfield  {author} {\bibinfo {author} {\bibfnamefont {D.}~\bibnamefont
  {Astefanesei}}, \bibinfo {author} {\bibfnamefont {D.}~\bibnamefont {Choque}},
  \bibinfo {author} {\bibfnamefont {J.}~\bibnamefont {Maggiolo}}, \ and\
  \bibinfo {author} {\bibfnamefont {R.}~\bibnamefont {Rojas}},\ }\href
  {\doibase 10.1103/PhysRevD.106.044032} {\bibfield  {journal} {\bibinfo
  {journal} {Phys. Rev. D}\ }\textbf {\bibinfo {volume} {106}},\ \bibinfo
  {pages} {044032} (\bibinfo {year} {2022})},\ \Eprint
  {http://arxiv.org/abs/2111.01337} {arXiv:2111.01337 [hep-th]} \BibitemShut
  {NoStop}%
\bibitem [{\citenamefont {Brown}\ and\ \citenamefont
  {York}(1993)}]{Brown:1992br}%
  \BibitemOpen
  \bibfield  {author} {\bibinfo {author} {\bibfnamefont {J.~D.}\ \bibnamefont
  {Brown}}\ and\ \bibinfo {author} {\bibfnamefont {J.~W.}\ \bibnamefont {York},
  \bibfnamefont {Jr.}},\ }\href {\doibase 10.1103/PhysRevD.47.1407} {\bibfield
  {journal} {\bibinfo  {journal} {Phys. Rev. D}\ }\textbf {\bibinfo {volume}
  {47}},\ \bibinfo {pages} {1407} (\bibinfo {year} {1993})},\ \Eprint
  {http://arxiv.org/abs/gr-qc/9209012} {arXiv:gr-qc/9209012} \BibitemShut
  {NoStop}%
\bibitem [{\citenamefont {Anabalon}(2012)}]{Anabalon:2012ta}%
  \BibitemOpen
  \bibfield  {author} {\bibinfo {author} {\bibfnamefont {A.}~\bibnamefont
  {Anabalon}},\ }\href {\doibase 10.1007/JHEP06(2012)127} {\bibfield  {journal}
  {\bibinfo  {journal} {JHEP}\ }\textbf {\bibinfo {volume} {06}},\ \bibinfo
  {pages} {127} (\bibinfo {year} {2012})},\ \Eprint
  {http://arxiv.org/abs/1204.2720} {arXiv:1204.2720 [hep-th]} \BibitemShut
  {NoStop}%
\bibitem [{\citenamefont {Anabalon}\ and\ \citenamefont
  {Oliva}(2012)}]{Anabalon:2012ih}%
  \BibitemOpen
  \bibfield  {author} {\bibinfo {author} {\bibfnamefont {A.}~\bibnamefont
  {Anabalon}}\ and\ \bibinfo {author} {\bibfnamefont {J.}~\bibnamefont
  {Oliva}},\ }\href {\doibase 10.1103/PhysRevD.86.107501} {\bibfield  {journal}
  {\bibinfo  {journal} {Phys. Rev. D}\ }\textbf {\bibinfo {volume} {86}},\
  \bibinfo {pages} {107501} (\bibinfo {year} {2012})},\ \Eprint
  {http://arxiv.org/abs/1205.6012} {arXiv:1205.6012 [gr-qc]} \BibitemShut
  {NoStop}%
\bibitem [{\citenamefont {Anabalon}\ \emph
  {et~al.}(2016{\natexlab{b}})\citenamefont {Anabalon}, \citenamefont
  {Astefanesei},\ and\ \citenamefont {Choque}}]{Anabalon:2016izw}%
  \BibitemOpen
  \bibfield  {author} {\bibinfo {author} {\bibfnamefont {A.}~\bibnamefont
  {Anabalon}}, \bibinfo {author} {\bibfnamefont {D.}~\bibnamefont
  {Astefanesei}}, \ and\ \bibinfo {author} {\bibfnamefont {D.}~\bibnamefont
  {Choque}},\ }\href {\doibase 10.1016/j.physletb.2016.08.049} {\bibfield
  {journal} {\bibinfo  {journal} {Phys. Lett. B}\ }\textbf {\bibinfo {volume}
  {762}},\ \bibinfo {pages} {80} (\bibinfo {year} {2016}{\natexlab{b}})},\
  \Eprint {http://arxiv.org/abs/1606.07870} {arXiv:1606.07870 [hep-th]}
  \BibitemShut {NoStop}%
\bibitem [{\citenamefont {Breitenlohner}\ and\ \citenamefont
  {Freedman}(1982)}]{Breitenlohner:1982bm}%
  \BibitemOpen
  \bibfield  {author} {\bibinfo {author} {\bibfnamefont {P.}~\bibnamefont
  {Breitenlohner}}\ and\ \bibinfo {author} {\bibfnamefont {D.~Z.}\ \bibnamefont
  {Freedman}},\ }\href {\doibase 10.1016/0370-2693(82)90643-8} {\bibfield
  {journal} {\bibinfo  {journal} {Phys. Lett. B}\ }\textbf {\bibinfo {volume}
  {115}},\ \bibinfo {pages} {197} (\bibinfo {year} {1982})}\BibitemShut
  {NoStop}%
\bibitem [{\citenamefont {Henneaux}\ \emph {et~al.}(2007)\citenamefont
  {Henneaux}, \citenamefont {Martinez}, \citenamefont {Troncoso},\ and\
  \citenamefont {Zanelli}}]{Henneaux:2006hk}%
  \BibitemOpen
  \bibfield  {author} {\bibinfo {author} {\bibfnamefont {M.}~\bibnamefont
  {Henneaux}}, \bibinfo {author} {\bibfnamefont {C.}~\bibnamefont {Martinez}},
  \bibinfo {author} {\bibfnamefont {R.}~\bibnamefont {Troncoso}}, \ and\
  \bibinfo {author} {\bibfnamefont {J.}~\bibnamefont {Zanelli}},\ }\href
  {\doibase 10.1016/j.aop.2006.05.002} {\bibfield  {journal} {\bibinfo
  {journal} {Annals Phys.}\ }\textbf {\bibinfo {volume} {322}},\ \bibinfo
  {pages} {824} (\bibinfo {year} {2007})},\ \Eprint
  {http://arxiv.org/abs/hep-th/0603185} {arXiv:hep-th/0603185} \BibitemShut
  {NoStop}%
\bibitem [{\citenamefont {Brown}\ and\ \citenamefont
  {Henneaux}(1986)}]{Brown:1986nw}%
  \BibitemOpen
  \bibfield  {author} {\bibinfo {author} {\bibfnamefont {J.~D.}\ \bibnamefont
  {Brown}}\ and\ \bibinfo {author} {\bibfnamefont {M.}~\bibnamefont
  {Henneaux}},\ }\href {\doibase 10.1007/BF01211590} {\bibfield  {journal}
  {\bibinfo  {journal} {Commun. Math. Phys.}\ }\textbf {\bibinfo {volume}
  {104}},\ \bibinfo {pages} {207} (\bibinfo {year} {1986})}\BibitemShut
  {NoStop}%
\bibitem [{\citenamefont {Aharony}\ \emph {et~al.}(2015)\citenamefont
  {Aharony}, \citenamefont {Gur-Ari},\ and\ \citenamefont
  {Klinghoffer}}]{Aharony:2015afa}%
  \BibitemOpen
  \bibfield  {author} {\bibinfo {author} {\bibfnamefont {O.}~\bibnamefont
  {Aharony}}, \bibinfo {author} {\bibfnamefont {G.}~\bibnamefont {Gur-Ari}}, \
  and\ \bibinfo {author} {\bibfnamefont {N.}~\bibnamefont {Klinghoffer}},\
  }\href {\doibase 10.1007/JHEP05(2015)031} {\bibfield  {journal} {\bibinfo
  {journal} {JHEP}\ }\textbf {\bibinfo {volume} {05}},\ \bibinfo {pages} {031}
  (\bibinfo {year} {2015})},\ \Eprint {http://arxiv.org/abs/1501.06664}
  {arXiv:1501.06664 [hep-th]} \BibitemShut {NoStop}%
\bibitem [{\citenamefont {Lola}(1998)}]{Lola:1998cr}%
  \BibitemOpen
  \bibfield  {author} {\bibinfo {author} {\bibfnamefont {S.}~\bibnamefont
  {Lola}},\ }\href {\doibase 10.22323/1.001.0059} {\bibfield  {journal}
  {\bibinfo  {journal} {PoS}\ }\textbf {\bibinfo {volume} {corfu98}},\ \bibinfo
  {pages} {059} (\bibinfo {year} {1998})},\ \Eprint
  {http://arxiv.org/abs/hep-ph/9903203} {arXiv:hep-ph/9903203} \BibitemShut
  {NoStop}%
\bibitem [{\citenamefont {Anabalon}(2014)}]{Anabalon:2012dw}%
  \BibitemOpen
  \bibfield  {author} {\bibinfo {author} {\bibfnamefont {A.}~\bibnamefont
  {Anabalon}},\ }\href {\doibase 10.1007/978-3-319-06761-2_1} {\bibfield
  {journal} {\bibinfo  {journal} {Springer Proc. Phys.}\ }\textbf {\bibinfo
  {volume} {157}},\ \bibinfo {pages} {3} (\bibinfo {year} {2014})},\ \Eprint
  {http://arxiv.org/abs/1211.2765} {arXiv:1211.2765 [gr-qc]} \BibitemShut
  {NoStop}%
\end{thebibliography}%

\end{document}